\documentclass[prepreint,twocolumn,trackchanges]{aastex62}
\usepackage{cprotect}
\hypersetup{linkcolor={blue},citecolor={blue}}
\received{\today}
\revised{\today}
\accepted{\today}
\submitjournal{ApJ}

\shorttitle{Interacting Supernovae in 2D}
\shortauthors{Suzuki, Moriya, \& Takiwaki}

\begin{document}
\title{Supernova ejecta interacting with a circumstellar disk. I. two-dimensional radiation-hydrodynamic simulations}

\correspondingauthor{Akihiro Suzuki (NAOJ fellow)}
\email{akihiro.suzuki@nao.ac.jp}

\author[0000-0002-7043-6112]{Akihiro Suzuki}
\affil{Division of Science, National Astronomical Observatory of Japan, 2-21-1 Osawa, Mitaka, Tokyo 181-8588, Japan}

\author[0000-0003-1169-1954]{Takashi J. Moriya}
\affil{Division of Science, National Astronomical Observatory of Japan, 2-21-1 Osawa, Mitaka, Tokyo 181-8588, Japan}
\affiliation{School of Physics and Astronomy, Faculty of Science, Monash University, Clayton, VIC 3800, Australia}

\author[0000-0003-0304-9283]{Tomoya Takiwaki}
\affil{Division of Science, National Astronomical Observatory of Japan, 2-21-1 Osawa, Mitaka, Tokyo 181-8588, Japan}
\affiliation{Center for Computational Astrophysics, National Astronomical Observatory of Japan, 2-21-1 Osawa, Mitaka, Tokyo 181-8588, Japan}



\begin{abstract}
We perform a series of two-dimensional radiation-hydrodynamic simulations of the collision between supernova ejecta and circumstellar media (CSM). 
The hydrodynamic interaction of a fast flow and the surrounding media efficiently dissipates the kinetic energy of the fast flow and considered as a dominant energy source for a specific class of core-collapse supernovae. 
Despite some observational evidence for aspherical ejecta and/or CSM structure, multi-dimensional effects in the ejecta-CSM interaction are relatively unexplored. 
Our numerical simulations equipped with an adaptive mesh refinement technique successfully reproduce hydrodynamic instabilities developing around the ejecta-CSM interface. 
We also investigate effects of disk-like CSM on the dynamical evolution of supernova ejecta and bolometric light curves. 
We find that emission powered by ejecta-disk interaction exhibits significant viewing angle dependence. 
For a line of sight close to the symmetry axis, the observer directly sees the supernova ejecta, leading to a short brightening timescale.  
For an observer seeing the emission through the CSM disk, thermal photons diffuse throughout the CSM and thus the light curve is severely smeared out. 
\end{abstract}



\section{Introduction}
It is firmly believed that the circum-stellar media (CSM) play important roles in stellar explosions. 
Massive stars are thought to produce CSMs by shedding a part of their envelopes via several mechanisms, e.g., stellar winds and binary interactions. 
When a massive star undergoes the gravitational collapse of the iron core, a core-collapse supernova (CCSN) happens and its bright emission illuminates the surrouding media. 
CCSNe showing observational signatures of hydrogen-rich CSMs are commonly found in SN surveys and they make up a special spectral class, known as type IIn SNe \citep{1990MNRAS.244..269S,1997ARA&A..35..309F,2017hsn..book..403S,2017hsn..book..843B}. 
Type IIn SNe are defined as SNe showing narrow emission/absorption line features superposed on commonly found broad P-Cygni profiles, thereby indicating the presence of slowly moving gas ahead of the fast SN ejecta. 
They constitute a non-negligible fraction of CCSNe ($\sim7\%$, e.g., \citealt{2011MNRAS.412.1441L,2011MNRAS.412.1522S,2017PASP..129e4201S}). 
The well-studied examples include SN 1988Z \citep{1991MNRAS.250..786S,1993MNRAS.262..128T,1993ApJ...419L..69V,1994MNRAS.268..173C,1999MNRAS.309..343A}, 1998S \citep{2000MNRAS.318.1093F,2001MNRAS.325..907F,2000ApJ...536..239L}, and 2010jl \citep{2011A&A...527L...6P,2011ApJ...730...34S,2012AJ....144..131Z,2014ApJ...797..118F,2014ApJ...781...42O}. 
The ejecta-CSM collision can efficiently dissipate the kinetic energy of the ejecta and release the dissipated energy as bright thermal radiation. 
Therefore, the ejecta-CSM interaction is considered as a dominant power source for luminous SNe like SN 2006gy \citep{2007ApJ...659L..13O,2007ApJ...666.1116S,2010ApJ...709..856S}.

Theoretical studies on the ejecta-CSM interaction have been conduced extensively in both analytical and numerical ways \citep[e.g.,][]{1982ApJ...258..790C,1982ApJ...259..302C,1994ApJ...420..268C,2007ApJ...671L..17S,2011ApJ...729L...6C,2011MNRAS.415..199M,2013MNRAS.428.1020M,2012ApJ...757..178G,2015MNRAS.449.4304D}, reaching the consensus that the required CSM masses can widely vary from an event to another. 
Most extreme cases of superluminous type IIn SNe often require as much as $\sim10\ M_\odot$. 
For such massive CSMs, the  star is covered by an optically thick media and the photosphere is no longer located at the stellar surface. 
The energy source of the bright optical emission, the ejecta--CSM interface, is initially hidden in the deep interior of the massive CSM. 
Thus, the energy produced due to the ejecta-CSM collision is reprocessed and then diffuses out through the photosphere in the CSM. 
More recent studies have shown that a considerable fraction of type II SNe also exhibit observational features indicating an enhanced CSM density in the immediate vicinity of exploding stars \citep{2014Natur.509..471G,2017NatPh..13..510Y,2018NatAs.tmp..122F}. 

Despite these growing evidence, how exactly massive stars shed their envelopes immediately before the gravitational collapse is still debated. 
For example, the most extreme case of SN 2006gy requires a mass-loss rate of the order of $\sim 1\ M_\odot$ yr$^{-1}$, which is obviously beyond normal rates inferred from Galactic massive stars \citep[e.g.,][]{2006A&A...457.1015H,2012A&A...540A.144S,2014ARA&A..52..487S} and cannot be explained by the current standard theory of steady stellar winds. 
Unusual mass-loss events associated with the final evolutionary state of massive stars, such as, wave-driven mass-loss \citep{2012MNRAS.423L..92Q,2014ApJ...780...96S,2017MNRAS.470.1642F,2018MNRAS.476.1853F}, envelope ejection due to binary interaction \citep[e.g.,][]{2012ApJ...752L...2C,2013ApJ...764L...6S}, fossil disks around massive stars \citep{2010MNRAS.409..284M}, and so on, may be responsible for the origin of massive CSMs. 
However, no general consensus has been reached yet. 
From the observational viewpoint, several Galactic luminous blue variables (LBVs; e.g., \citealt{1994PASP..106.1025H}), e.g., $\eta$ Carinae, have massive CSMs in their vicinity. 
Such mass-loss events long before the iron core-collapse are likely to be eruptive \citep[e.g.,][]{2007Natur.447..829P}, thereby giving rise to optical transients associated with non-terminal explosions, i.e., SN imposters \citep{2000PASP..112.1532V}, such as SN 2009ip \citep{2010AJ....139.1451S,2011ApJ...732...32F,2013MNRAS.430.1801M,2013ApJ...767....1P}, characterized by less bright and redder optical emission than SNe. 

One of the highly uncertain but important issues among interacting SNe is multi-dimensional effects. 
Deviations from spherical symmetry are expected even in cases where ejecta and CSM are both nearly spherical because of the development of (radiation-)hydrodynamic instabilities, the Rayleigh-Taylor instability and the Vishniac instability \citep{1983ApJ...274..152V,1987ApJ...313..820R}, around the ejecta-CSM interface. 
In addition, the CSM itself can be asymmetric and/or clumpy \citep[e.g.,][]{1994MNRAS.268..173C}. 

Currently, there is growing observational evidence for SNe interacting with aspherical CSMs.
In this paper we especially focus on SN ejecta interacting with disk-like CSMs. 
The well-known example is SN 1987A in the Large Magellanic cloud, which is a historical SN interacting with a spatially resolved torus-like structure \citep[e.g.,][]{1995ApJ...452..680B,1995ApJ...439..730P,2016ApJ...833..147L,2016ARA&A..54...19M}, although its origin is still debated. 
For extragalactic SNe, aspherical ejecta and CSM strucutre can be probed by a number of observational imprints, such as, polarization signals, emission lines with intermediate widths, and asymmetric nebular line profiles \citep[e.g.,][]{1982ApJ...263..902S,1991A&A...246..481H,2001ApJ...550.1030W,2008ARA&A..46..433W,2011A&A...527L...6P}. 
Some type IIn SNe are indeed suspected to have aspherical CSMs. 
For example, SN 1998S exhibited a high degree of linear polarization and emission lines with asymmetric profiles \citep{2000ApJ...536..239L}. 
\cite{2012ApJ...756..173S} argued that the type IIn SN 2006dj likely experienced eruptive mass-loss events producing highly anisotropic CSMs. 
In the case of the 2012 eruptive event of SN 2009ip, although it is still debated whether it was an LBV eruption or the terminal explosion of a massive star, the spectroscopic and spectropolarimetric studies suggested that the ejected material was interacting with a disk-like medium \citep{2013MNRAS.433.1312F,2014AJ....147...23L,2014ApJ...780...21M,2014MNRAS.442.1166M,2014MNRAS.438.1191S}. 
\cite{2014ApJ...780..184K,2016ApJ...832..194K} analyzed X-ray spectra of SN 2005kd, 2006dj and 2010jl, and pointed out that the two distinguishing emission components can be interpreted as emission absorbed by a torus-like CSM. 
The number of type IIn SNe with detailed follow-up observations is rapidly increasing thanks to modern transient surveys, e.g., PTF11iqb \citep{2015MNRAS.449.1876S}, SN 2012ab \citep{2018MNRAS.475.1104B}, and PTF12glz \citep{2018arXiv180804232S}, and we can expect more in the coming future. 
Very recently, \cite{2019arXiv190605812N} present type IIn SN samples from the untargeted PTF survey and distributions of some observational properties, such as, the rise and decline times and the peak luminosity. 

Despite its potential importance, the dynamical evolution of SN ejecta with an aspherical CSM is relatively poorly investigated compared to spherical counterparts \citep{1996ApJ...472..257B,2010MNRAS.407.2305V,2016MNRAS.458.1253V,2018ApJ...856...29M,2019A&A...625A..24K}. 
\cite{2016MNRAS.458.1253V} performed 2D radiation-hydrodynamic simulations of SN ejecta-CSM collisions in various settings including spherical/aspherical ejecta and CSM.
Although they demonstrate that aspherical CSMs do affect light curves of the emission powered by the ejecta-CSM interaction, the spatial resolution of their numerical simulations was not enough to resolve hydrodynamic instabilities expected in the ejecta-CSM interface. 
More recently, \cite{2018ApJ...856...29M} performed 2D hydrodynamic simulations of SN ejecta interacting with a CSM disk and evaluate the energy dissipation rate of the kinetic energy of the SN ejecta, which is then used to estimate the luminosity of the ejecta powered by the CSM interaction.
Their light curve modeling, however, is based on standard one-zone model \citep{1982ApJ...253..785A,1996snih.book.....A,2012ApJ...746..121C} and the viewing angle effect is not investigated. 
\cite{2019A&A...625A..24K} also performed a series of 2D hydrodynamic simulations of SN ejecta-CSM disk interaction. 
They focus on the dynamical evolution of the SN ejecta in the presence of a CSM disk. 

In this work, we investigate multi-dimensional effects in SN ejecta-CSM interaction by performing 2D radiation-hydrodynamic simulations. 
We particularly focus on SN ejecta interacting with CSM disks with different masses and opening angles. 
In Section \ref{sec:simulation_setups}, we describe our numerical code and setups. 
In Section \ref{sec:results}, we present the results of the simulations and compare them with one-dimensional spherical simulations. 
In addition, we obtain the approximate color temperature evolution by locating the photosphere (Section \ref{sec:post_process}). 
We discuss observational implications in Section \ref{sec:discussion}. 
Finally, Section \ref{sec:conclusions} concludes this paper. 
Appendices \ref{sec:numerical_integration_detail} and \ref{sec:locating_photosphere} provide numerical procedures in details. 
We adopt the unit $c=1$ unless otherwise noted. 

\section{Simulation setups}\label{sec:simulation_setups}
We perform two-dimensional special relativistic radiation-hydrodynamic simulations of SN ejecta colliding with a CSM. 
The numerical code have been developed by one of the authors and applied to a two-dimensional study on aspherical SN shock breakout \citep{2016ApJ...825...92S}. 
Our treatment of radiative transfer is based on the so-called two-temperature approximation. 
Although this simple treatment is not always appropriate, it is a convenient and widely used first step toward extending radiation-hydrodynamic simulations in multi-dimension. 
The code employs an adaptive mesh refinement technique \citep[AMR;][]{1989JCoPh..82...64B} so that the ejecta-CSM interface is well resolved, while the numerical domain covers the whole SN ejecta. 
In this section, we describe the governing equations and the simulation setups. 

\subsection{Equations of radiation-hydrodynamics\label{sec:equations_of_radiation_hydrodynamics}}
We solve the following equations for radiation-hydrodynamics. 
In this work, we perform simulations in 1D spherical and 2D cylindrical coordinates. 
In the following, we present equations for 3D cartesian coordinates for the purpose of keeping them general. 
Thus, the indices $i$ and $j$ run from $1$ to $3$ and the usual summation convention is used unless otherwise noted. 
When applying the following method to a specific curvilinear coordinate system, some geometrical factors should be introduced correctly. 

The numerical code solves the temporal evolutions of the frequency-integrated radiation energy density $E_\mathrm{r}$ and the flux $F_\mathrm{r}^i$,
\begin{equation}
\frac{\partial E_\mathrm{r}}{\partial t}+\frac{\partial F_\mathrm{r}^i}{\partial x^i}=G^0,
\label{eq:eq_Erad}
\end{equation}
and
\begin{equation}
\frac{\partial F_\mathrm{r}^i}{\partial t}+\frac{\partial P_\mathrm{r}^{ij}}{\partial x^{j}}=G^i,
\label{eq:eq_flux}
\end{equation}
where $P_\mathrm{r}^{ij}$ is the radiation pressure tensor. 
We note that the variables $E_\mathrm{r}$, $F_\mathrm{r}$, and $P_\mathrm{r}^{ij}$ in these equations are defined in the laboratory frame \citep[see, e.g.,][]{1984oup..book.....M}. 
The right-hand sides of these equations represent the changes in the radiation energy density and the radiative flux per unit time caused by absorption, emission, and scattering of photons (see, Section \ref{sec:source_term}). 
These equations are coupled with the hydrodynamic equations for the density $\bar{\rho}$, the velocity $\beta^i$, and the gas energy density $\bar{E}_\mathrm{g}$,
\begin{equation}
\frac{\partial (\bar{\rho}\Gamma)}{\partial t}+\frac{\partial (\bar{\rho}\Gamma\beta^i)}{\partial x^i}=0,
\label{eq:eq_continuity}
\end{equation}
\begin{eqnarray}
&&\frac{\partial [(\bar{\rho}+\bar{E}_\mathrm{g}+\bar{P}_\mathrm{g})\Gamma^2\beta^i]}{\partial t}
\nonumber\\
&&\hspace{2em}+\frac{\partial \left[(\bar{\rho}+\bar{E}_\mathrm{g}+\bar{P}_\mathrm{g})\Gamma^2\beta^i\beta^j+\bar{P}_\mathrm{g}\right])}{\partial x^j}=-G^i,
\label{eq:eq_momentum}
\end{eqnarray}
and
\begin{eqnarray}
&&\frac{\partial [(\bar{\rho}+\bar{E}_\mathrm{g}+\bar{P}_\mathrm{g})\Gamma^2-\bar{P}_\mathrm{g}]}{\partial t}
\nonumber\\
&&\hspace{4em}
+\frac{\partial [(\bar{\rho}+\bar{E}_\mathrm{g}+\bar{P}_\mathrm{g})\Gamma^2\beta^i]}{\partial x^i}=-G^0,
\label{eq:eq_Egas}
\end{eqnarray}
where $\bar{P}_\mathrm{g}$ is the gas pressure and $\Gamma=(1-\beta^2)^{-1/2}$ is the Lorentz factor. 
We note that physical quantities defined in the comoving frame are expressed by letters with overbars, e.g., $\bar{Q}$. 
We assume an ideal gas equation of state with an adiabatic index $\gamma$, 
\begin{equation}
\bar{P}_\mathrm{g}=(\gamma-1)\bar{E}_\mathrm{g},
\label{eq:eos}
\end{equation}
and assume $\gamma=5/3$. 
The gas temperature $\bar{T}_\mathrm{g}$ is defined as
\begin{equation}
\bar{T}_\mathrm{g}=\frac{\mu m_\mathrm{u}(\gamma-1)\bar{E}_\mathrm{g}}{\bar{\rho} k_\mathrm{B}},
\end{equation}
where $k_\mathrm{B}$ and $m_\mathrm{u}$ are the Boltzmann constant and the atomic mass unit. 
The mean molecular weight $\mu$ is approximately calculated by using the hydrogen and helium mass fractions, $X_\mathrm{h}$ and $X_\mathrm{he}$, in the following way,
\begin{equation}
\mu^{-1}=2X_\mathrm{h}+0.75X_\mathrm{he}+0.56(1-X_\mathrm{h}-X_\mathrm{he}).
\end{equation}
In the following simulations, the gas temperature is mostly lower than $\bar{T}_\mathrm{g}<5\times 10^9$ K, above which electrons behave as a relativistic gas (i.e., the adiabatic index close to $4/3$). 
Therefore, the assumption of an electron gas in the non-relativistic regime $(\gamma=5/3)$ is justified. 

\subsubsection{Advection terms}
The left hand sides of Equations (\ref{eq:eq_Erad})--(\ref{eq:eq_Egas}) represent the transport of mass, momentum, and energy in the physical space. 
Therefore, these equations with the source terms being zero can be integrated in standard numerical ways. 
We adopt the same steps as the previous work \citep{2016ApJ...825...92S}. 
For Equations (\ref{eq:eq_Erad}) and (\ref{eq:eq_flux}), we employ the so-called M1 closure scheme \citep{1984JQSRT..31..149L}, where the Eddington tensor $D^{ij}=P_\mathrm{r}^{ij}/E_\mathrm{r}$ is given by the following analytic function of $E_\mathrm{r}$ and $F^i_\mathrm{r}$;
\begin{equation}
D^{ij}=\frac{1-\chi}{2}\delta^{ij}+\frac{3\chi-1}{2}n^in^j,
\end{equation} 
where $n^i=F_\mathrm{r}^i/|F_\mathrm{r}|$ and the parameter $\chi$ is calculated as follows,
\begin{equation}
\chi=\frac{3+4f^if_i}{5+2\sqrt{4-3f^if_i}},
\end{equation}
with $f^i=F_\mathrm{r}^i/E_\mathrm{r}$. 
For the spatial reconstruction of $E_\mathrm{r}$ and $F^i_\mathrm{r}$, we employ the 3rd-order weighted essentially non-oscillatory scheme \citep[WENO;][]{1994JCoPh.115..200L,1996JCoPh.126..202J}. 

For the hydrodynamics equations, we also use the commonly adopted Harten-Lax-van Leer-Einfeldt (HLLE) Riemann solver \citep{2005MNRAS.364..126M} combined with the 3rd-order MUSCL reconstruction.

\subsubsection{Source terms\label{sec:source_term}}
The equations for radiation field and hydrodynamic variables are related by the coupling terms $G^0$ and $G^i$, whose functional forms are as follows,
\begin{equation}
G^0=\Gamma\bar{\rho}\bar{\kappa_\mathrm{a}}(a_\mathrm{r}\bar{T}_\mathrm{g}^4-\bar{E}_\mathrm{r})
-\Gamma\bar{\rho}(\bar{\kappa}_\mathrm{a}+\bar{\kappa}_\mathrm{s})\beta_i\bar{F}_\mathrm{r}^i,
\label{eq:G0}
\end{equation}
and
\begin{eqnarray}
G^i&=&
-\bar{\rho}(\bar{\kappa}_\mathrm{a}+\bar{\kappa}_\mathrm{s})\bar{F}_\mathrm{r}^i
+\Gamma\bar{\rho}\bar{\kappa}_\mathrm{a}\left(a_\mathrm{r}\bar{T}_\mathrm{g}^4-\bar{E}_\mathrm{r}\right)\beta^i
\nonumber\\&&
-\frac{\Gamma^2}{\Gamma+1}\bar{\rho}(\bar{\kappa}_\mathrm{a}+\bar{\kappa}_\mathrm{s})\beta^i\beta_j\bar{F}_\mathrm{r}^j,
\label{eq:Gi}
\end{eqnarray}
where $\bar{\kappa}_\mathrm{a}$ and $\bar{\kappa}_\mathrm{s}$ are absorption and scattering opacities, respectively. 
We have assumed that the scattering process is isotropic in the comoving frame. 
The radiation energy density and the flux appearing in these expressions are defined in the comoving frame and are related with those in the laboratory frame through the Lorentz transformation; 
\begin{equation}
{E}_\mathrm{r}=\Gamma^2\left(\bar{E}_\mathrm{r}+2\beta_i\bar{F}_\mathrm{r}^i+\beta_i\beta_j\bar{P}_\mathrm{r}^{jk}\right),
\label{eq:transformation_Er}
\end{equation}
and
\begin{eqnarray}
\hspace{-2em}
{F}_\mathrm{r}^i&=&\Gamma \bar{F}_\mathrm{r}^i+\Gamma\beta_j \bar{P}_\mathrm{r}^{ij}
\nonumber\\
&&+\Gamma^2\beta^i
\left(\bar{E}_\mathrm{r}+\frac{2\Gamma+1}{\Gamma+1}\beta_j\bar{F}_\mathrm{r}^j+\frac{\Gamma}{\Gamma+1}\beta_j\beta_k\bar{P}_\mathrm{r}^{jk}\right),
\label{eq:transformation_Fr}
\end{eqnarray}
\citep[see, e.g.,][]{1984oup..book.....M}, which close the equations. 
The numerical methods to solve these equations are presented in Appendix \ref{sec:source_term_numerical}. 

Our treatment of electron scattering has some limitations. 
While the free-free process immediately realizes gas-radiation equilibrium in sufficiently dense material, it is not always effective. 
For shocks propagating in dilute gas, for example, post-shock gas density could be so small that the free-free process is not effective for creating an enough number of photons for the equilibrium. 
Alternatively, Compton scattering can be a dominant energy transfer process for gas and radiation. 
With an insufficient number of photons, the photon spectrum becomes a Wien function rather than a Planck function.  
This circumstance is expected for SN shock breakout in a dilute stellar wind  \citep{1976ApJS...32..233W}. 
In this way, our two-temperature treatment sometimes do not capture the gas-radiation coupling correctly. 
In this study, however, we focus on shock breakout from a dense CSM, where free-free process produce enough photons to achieve bright thermal emission. 

\subsubsection{Opacity}
We consider free-free absorption and electron scattering as the dominant radiative processes in this setting. 
We assume the commonly adopted opacity formulae,
\begin{equation}
\bar{\kappa}_\mathrm{a}=3.7\times 10^{22}(1+X_\mathrm{h})(X_\mathrm{h}+X_\mathrm{he})\rho \bar{T}_\mathrm{g}^{-7/2}\ \mathrm{cm^2\ g^{-1}},
\end{equation}
for free-free absorption, and
\begin{equation}
\bar{\kappa}_\mathrm{s}=0.2(1+X_\mathrm{h})\ \mathrm{cm^2\ g^{-1}},
\label{eq:kappa_es}
\end{equation}
for electron scattering (in cgs units, see, e.g., \citealt{1979rpa..book.....R}). 
Here, we assume a fully ionized gas. 
We note that this is not always a valid assumption. 
As we shall see below, the photospheric temperature decreases down to the recombination temperature of hydrogen $<6000$ K at later epochs. 
Then, the assumption of fully ionized gas is no longer justified.

\subsection{Initial and boundary conditions}
For simulations in the 2D cylindrical coordinates $(r,z)$, the numerical domain covers the region with $0\leq r\leq 6.4\times 10^{16}$ cm and $-6.4\times 10^{16}$ cm$\leq z\leq6.4\times10^{16}$ cm. 
A reflection boundary condition is imposed at the symmetry axis $r=0$, while free boundary conditions are imposed for the other boundaries.

\subsubsection{SN ejecta}
We initially assume freely expanding spherical supernova ejecta. 
We start our simulation at $t_0=1000$ s. 
Thus, the radial velocity of a layer at a radius $R$ is initially given by 
\begin{equation}
v=\frac{R}{t_0},
\end{equation}
for $v<v_\mathrm{max}$ and $v=0$ otherwise. 
Here we denote the three-dimensional radius by $R\equiv (r^2+z^2)^{1/2}$, in order to distinguish it from the radial coordinate $r$. 
We also introduce the inclination angle $\theta=\cos^{-1}(r/R)$ to specify an angle measured from the symmetry axis $z=0$. 
The initial density structure is assumed to be the commonly adopted broken power-law distribution with the break velocity $v_\mathrm{br}$ \citep{1989ApJ...341..867C,1999ApJ...510..379M}, 
\begin{equation}
\rho_\mathrm{ej}(r,z)=\frac{f_3M_\mathrm{ej}}{4\pi v_\mathrm{br}^3t_0^3}g(R/t_0),
\label{eq:rho_sn}
\end{equation}
for $R<v_\mathrm{max}t_0$, where $M_\mathrm{ej}$ is the total mass of the ejecta and 
\begin{equation}
f_l=\frac{(m-l)(l-\delta)}{m-\delta-(l-\delta)(v_\mathrm{br}/v_\mathrm{max})^{m-l}}.
\end{equation}
For a sufficently steep outer density slope and a large $v_\mathrm{max}/v_\mathrm{br}$, this expression is approximated as follows,
\begin{equation}
f_l\simeq \frac{(m-l)(l-\delta)}{m-\delta}.
\end{equation}
The non-dimensional function $g(v)$ is given by
\begin{equation}
g(v)=\left\{
\begin{array}{ccl}
\left(\frac{v}{v_\mathrm{br}}\right)^{-\delta}&\mathrm{for}&v\leq v_\mathrm{br},\\
\left(\frac{v}{v_\mathrm{br}}\right)^{-m}&\mathrm{for}&v_\mathrm{br}<v.\\
\end{array}
\right.
\end{equation}
The exponents $\delta$ and $m$ are fixed to be $\delta=1$, and $m=10$ throughout this study. 
The break velocity $v_\mathrm{br}$ is determined by specifying the total mass and kinetic energy, $M_\mathrm{ej}$ and $E_\mathrm{sn}$, of the ejecta,
\begin{equation}
v_\mathrm{br}=\left(\frac{2f_5E_\mathrm{sn}}{f_3M_\mathrm{ej}}\right)^{1/2}
\simeq \left[\frac{2(m-5)(5-\delta)E_\mathrm{sn}}{(m-3)(3-\delta)M_\mathrm{ej}}\right]^{1/2}.
\end{equation}

We assume that the initial gas internal energy density of the ejecta is proportional to the initial kinetic energy distribution,
\begin{equation}
\bar{E}_\mathrm{g}(r,z)=\epsilon \frac{\rho_\mathrm{ej}(r,z)}{2}\left(\frac{R}{t_0}\right)^2,
\end{equation}
with $\epsilon=0.05$. 
The internal to kinetic energy ratio is expected to be around unity at the very beginning of the expansion of the SN ejecta. 
Then, the internal energy rapidly decreases in the absence of any heating source and immediately becomes negligible compared to the kinetic counterpart. 
The expanding outer SN envelop with a power-law radial density profile is obtained as an asymptotic behavior after the internal energy becomes negligible to the kinetic energy \citep{1989ApJ...341..867C}. 
Therefore, from the beginning, we set the internal energy of the ejecta to only $5\%$ of the kinetic one so that the radial density structure mimic the cold ejecta realized well after the explosion. 
This treatment is justified when the initial thermal energy loaded on the SN ejecta less likely contributes to the bolometric luminosity as in interacting SNe. 
The radiation energy density and the radiative flux are initially zero. 
Although there is no radiation field in the ejecta at the beginning of the simulation, the thermal equilibrium between gas and radiation, where the gas and radiation temperatures are identical, is immediately achieved in the SN ejecta. 

In this study, we fix the total mass and the initial kinetic energy of the ejecta to be $M_\mathrm{ej}=10M_\odot$ and $E_\mathrm{sn}=10^{51}$ erg. 
Thus, the corresponding break velocity is $v_\mathrm{br}=3.8\times 10^3$ km s$^{-1}$. The outer ejecta initially extend out to $v_\mathrm{max}t_0$ at time $t_0$, where the outermost layer is adjacent to the inner edge of the CSM. 
In the following, we set $v_\mathrm{max}=1.26\times 10^9$ cm s$^{-1}$ or $v_\mathrm{max}t_0=1.26\times 10^{12}$ cm, which corresponds to $v_\mathrm{br}/v_\mathrm{max}=0.3$.

\subsubsection{CSM}
\begin{figure*}
\begin{center}
\includegraphics[scale=0.28]{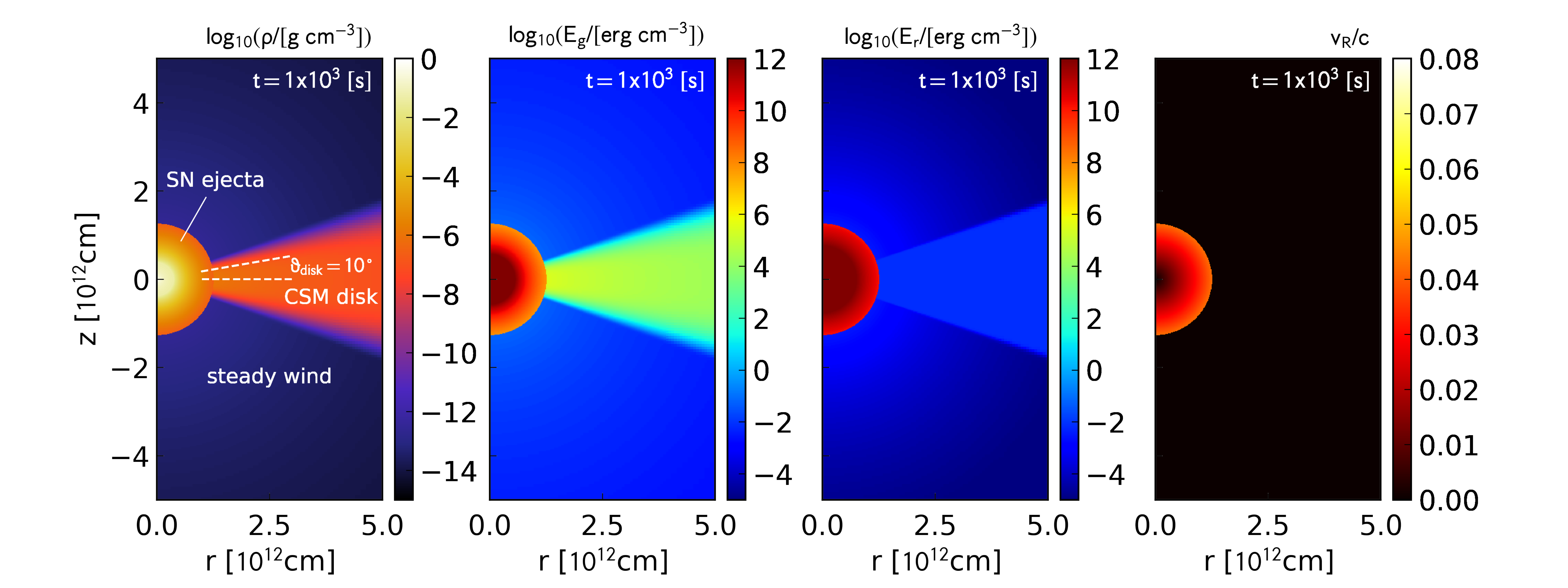}
\cprotect\caption{Spatial distributions of the density, the gas and radiation energy densities, and the radial velocity (from left to right) shortly after $t=t_0$ for model \verb|D10_M10|.}
\label{fig:initial}
\end{center}
\end{figure*}

We consider a centrally concentrated CSM embedded in a steady wind. 
Such CSM are thought to be an outcome of still unclear mass-loss activities of massive stars 10--1000 years before their core-collapse and therefore their density and velocity structures are highly uncertain. 
We assume that both the CSM and wind densities follow an inverse square law. 
The dense CSM is truncated at a radius of $5\times 10^{15}$ cm, which corresponds to an enhanced mass-loss 10--100 years prior to the core-collapse for a wind velocity of 10--100 km s$^{-1}$ (carbon- and oxygen-burning stages). 
Dense CSMs confined within a few $10^{15}$ cm are often adopted in light curve modellings of type IIn SN \citep[e.g.,][]{2012ApJ...757..178G,2013MNRAS.428.1020M}.  

Thus, for spherical CSMs, the density profile is described by
\begin{equation}
\rho(r,z)=\rho_\mathrm{wind}(R)+\rho_\mathrm{sph}(R),
\label{eq:rho_csm}
\end{equation}
where the wind and CSM components are given by
\begin{equation}
\rho_\mathrm{wind}(R)
=A_\star R^{-2},
\end{equation}
and
\begin{equation}
\rho_\mathrm{sph}(R)=\frac{A_\mathrm{sph}}{R^{2}}\exp\left[-\left(\frac{R}{R_\mathrm{csm}}\right)^{p}\right],
\end{equation}
with $p=10$. 
In the following, the wind density parameter is set to $A_\star=5\times 10^{11}$ g cm$^{-1}$, which corresponds to a steady mass-loss at a rate of $10^{-6}\ M_\odot$ yr$^{-1}$ for a constant wind velocity of $100$ km s $^{-1}$. 
This value leads to a sufficiently dilute wind component so that it does not affect the expansion of the SN ejecta. 
The exponential factor in this spherical CSM density profile realizes a smooth cutoff around $R=R_\mathrm{csm}$, beyond which the wind component dominates. 
The CSM mass $M_\mathrm{csm}$ is expressed as a function of the normalization constant  $A_\mathrm{sph}$ and the outer radius $R_\mathrm{csm}$,
\begin{eqnarray}
M_\mathrm{csm}&=&4\pi \int_0^{\infty} \rho_\mathrm{sph}(R)R^2dR
\nonumber\\
&=&4\pi A_\mathrm{sph}R_\mathrm{csm}\Gamma(1+1/p),
\end{eqnarray}
where $\Gamma(x)$ is a Gamma function and $\Gamma(1+1/p)\simeq 0.951$ for $p=10$.

For a CSM disk, we assume the following functional form, 
\begin{equation}
\rho(r,z)=\rho_\mathrm{wind}(R)+\rho_\mathrm{disk}(r,z),
\label{eq:rho_disk}
\end{equation}
with
\begin{equation}
\rho_\mathrm{disk}(r,z)=\frac{A_\mathrm{disk}}{R^2}\exp\left[-\left(\frac{R}{R_\mathrm{csm}}\right)^{p}-\left(\frac{\vartheta}{\vartheta_\mathrm{disk}}\right)^{q}\right],
\end{equation}
where $\vartheta$ is an angle measured from the equator, 
\begin{equation}
\vartheta=\cos^{-1}\left(\frac{r}{R}\right).
\end{equation}
Throught this work, we assume $q=4$. 
Therefore, the density is enhanced around the region with a half-opening angle of $\vartheta_\mathrm{disk}$. 
The parameter $q$ determines the slope of the upper and lower edges of the CSM disk. 
Although we adopt a very simplified CSM disk model, there would be room for improvement for future studies. 
For example, by assuming a steady disk, the scale height of the upper and lower edges of the disk is determined by the disk temperature. 
The geometry and structure of CSM disks would offer us a hint toward how they are produced and thus should be studied in more detail. 
The CSM mass $M_\mathrm{csm}$ is expressed as a function of $R_\mathrm{csm}$ and $\vartheta_\mathrm{disk}$,
\begin{equation}
M_\mathrm{csm}=4\pi A_\mathrm{disk}\Gamma(1+1/p)F_{q}(\vartheta_\mathrm{disk})R_\mathrm{csm},
\end{equation}
where $F_q(\vartheta_\mathrm{disk})$ is the covering fraction of the disk with respect to the solid angle,
\begin{equation}
F_q(\vartheta_\mathrm{disk})=\int_0^{\pi/2}
\exp\left[-\left(\frac{\vartheta}{\vartheta_\mathrm{disk}}\right)^q\right]\cos\vartheta d\vartheta.
\label{eq:F_q}
\end{equation}
The covering fraction yields $F_{4}(\vartheta_\mathrm{disk})=0.157$, and $0.310$, for $\vartheta_\mathrm{disk}=10^\circ$ and $20^\circ$. In Figure \ref{fig:initial}, we show the spatial distributions of the density, gas and radiation energy densities, and the radial velocity shortly after $t=t_0$ for model \verb|D10_M10|(see, Table \ref{table:models} for the model parameters).


\subsection{Adaptive mesh refinement}
The numerical code is equipped with an AMR technique  for better capturing discontinuities and tiny structures developing in the ejecta-CSM interface. 
The numerical domain is covered by $512\times 1024$ numerical cells at the lowest AMR level. 
Then, specific regions are successively covered by numerical cells with finer spatial resolutions. 
The domain with the coarsest resolution is referred to as AMR level $0$ and an increase in the AMR level by $1$ corresponds to a finer resolution by a factor of $2$. 
The highest AMR level is initially set to $l=13$, at which $2^{13}$ times finer resolution than the lowest AMR level is realized. 
This results in a minimum resolved length of $1.5\times10^{10}$ cm at $l=13$. 

Since the SN ejecta are expanding with time, a larger number of numerical cells are required at later epochs. 
Thus, we decrease the highest AMR level one by one as the ejecta expand and it reaches $l=4$ at the end of the simulation. 
Although the smallest cell size increases with time as a result of this coarsening, the minimum resolved length is guaranteed to be small compared to the physical scale of the expanding ejecta. 
In this way, we follow the evolution of the ejecta-CSM interaction up to $t\simeq 200$ days.

\subsection{Light curve calculation}
One of the properties of the radiation field that we can directly obtain from a single simulation is the bolometric light curve. 
For the light curve calculation, we consider an outgoing radiative flux at a distance, $R_\mathrm{obs}$, from the center of the ejecta. 
We define the viewing angle $\Theta_\mathrm{obs}$ measured from the symmetry axis $r=0$ and record the temporal evolution of the projected radiative flux, 
\begin{equation}
F_\mathrm{out}=l_{\mathrm{v},i}F_\mathrm{r}^i,
\label{eq:f_out}
\end{equation}
at the coordinates $(r,z)=(R_\mathrm{obs}\sin\Theta_\mathrm{obs},R_\mathrm{obs}\sin\Theta_\mathrm{obs})$.
The direction vector $l_\mathrm{v}^i$ is given by
\begin{equation}
l_\mathrm{v}^i=(\sin\Theta_\mathrm{obs},\cos\Theta_\mathrm{obs}).
\end{equation}
In other words, the radiative flux is projected on to the line of sight $l_\mathrm{v}^i$. 

The isotropic equivalent bolometric luminosity is simply calculated by using this outgoing flux,
\begin{equation}
L_\mathrm{bol,iso}(t,\Theta_\mathrm{obs})=4\pi R_\mathrm{obs}^2F_\mathrm{out}.
\end{equation}
We also define the cumulative radiated energy, $E_\mathrm{rad,iso}(t,\Theta_\mathrm{obs})$, by integrating the bolometric luminosity up to $t$,
\begin{equation}
E_\mathrm{rad,iso}(t,\Theta_\mathrm{obs})=\int^t_{t_0}L_\mathrm{bol,iso}(t',\Theta_\mathrm{obs})dt'.
\label{eq:E_rad}
\end{equation}
In the following simulations, we set the distance of the observer to be $R_\mathrm{obs}=5\times 10^{16}$ cm.

\begin{figure}
\begin{center}
\includegraphics[scale=0.5]{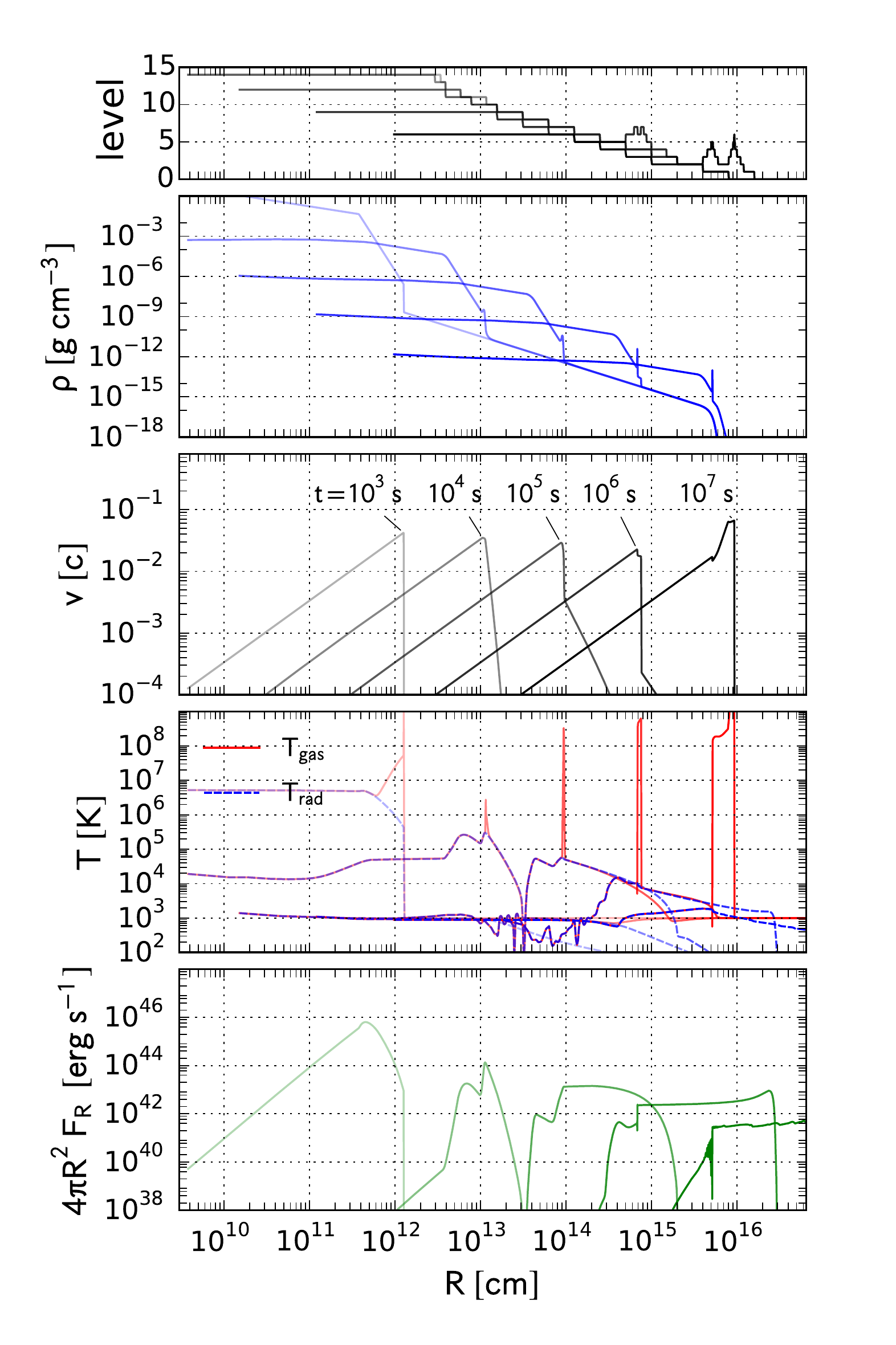}
\caption{Results of the 1D spherical simulation with $M_\mathrm{csm}=0.1M_\odot$. 
The radial profiles of the AMR level, density, radial velocity, gas and radiation temperatures, and the luminosity are plotted from top to bottom. 
The profiles at $t=10^3$, $10^4$, $10^5$, $10^6$, and $10^7$ s are shown in each panel. }
\label{fig:radial_1d_m01}
\end{center}
\end{figure}

\begin{figure}
\begin{center}
\includegraphics[scale=0.50]{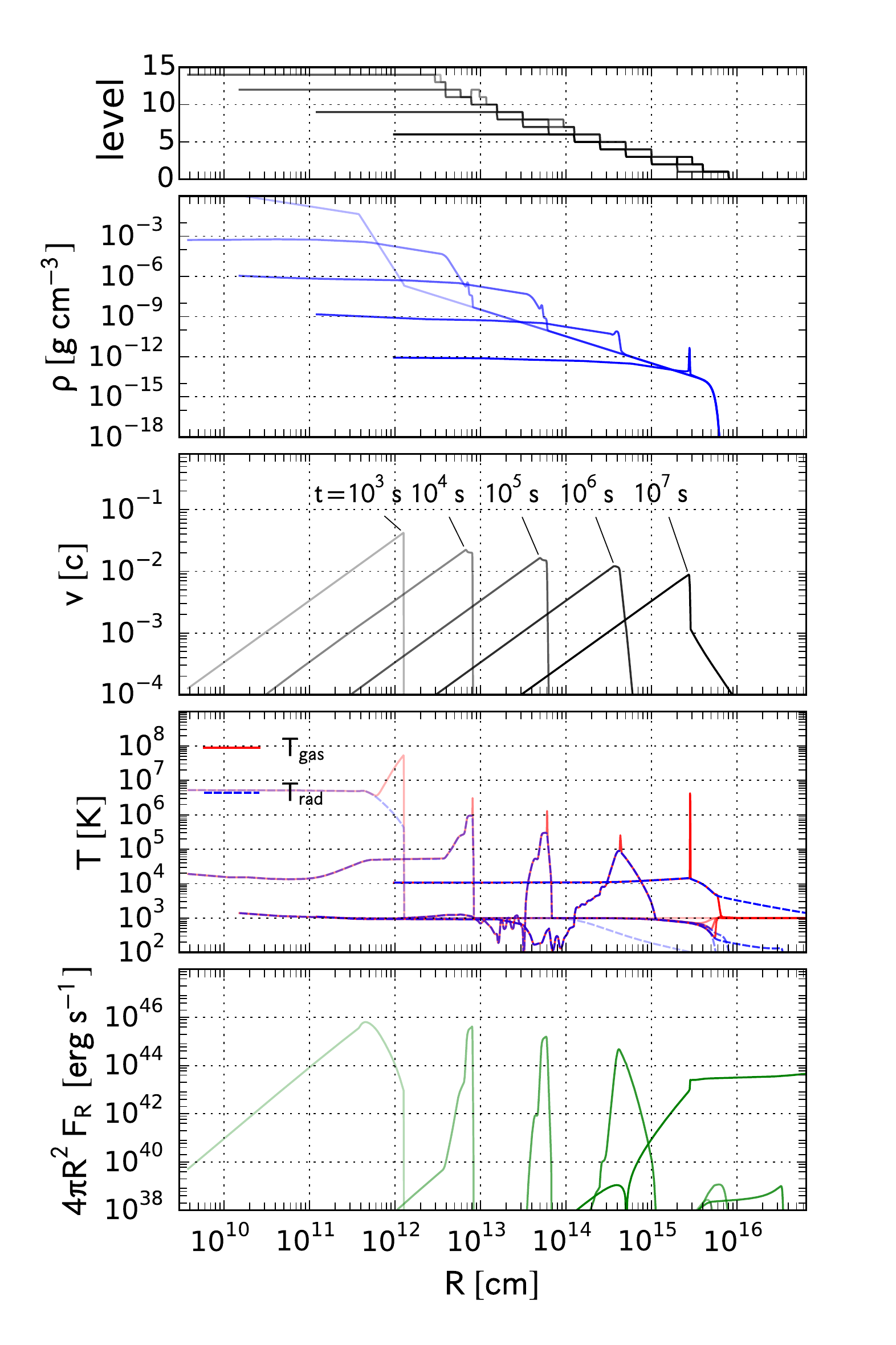}
\caption{Same as Figure \ref{fig:radial_1d_m01}, but for $M_\mathrm{csm}=10M_\odot$ }
\label{fig:radial_1d_m10}
\end{center}
\end{figure}

\section{Simulation results\label{sec:results}}
We performed numerical simulations of the ejecta-CSM interaction in 1D spherical and 2D cylindrical coordinates.
The 1D and 2D results are presented in Sections \ref{sec:1d_spherical_models} and \ref{sec:spherical_csm}, respectively.
The compositions of the ejecta and the CSM are assumed to be $X_\mathrm{h}=0.73$ and $X_\mathrm{he}=0.25$ in the whole numerical domain.

We carry out 2D simulations with spherical CSMs and disk-like CSMs with two different half opening angles of $10^\circ$ and $20^\circ$.  
The results are presented in Section \ref{sec:disk_csm}. 
For each configuration, we assume three different CSM masses, $M_\mathrm{csm}=0.1$, $1.0$, and $10\ M_\odot$, while the outer radius of the CSM is fixed to be $R_\mathrm{csm}=5\times 10^{15}$ cm. 
In Table \ref{table:models}, we provide the CSM parameters for the 9 models.

\begin{table}
\begin{center}
  \caption{Model descriptions}
\begin{tabular}{lrrr}
\hline\hline\\
Name&$M_\mathrm{csm}[M_\odot]$&$\vartheta_\mathrm{disk}[\mathrm{deg}]$&$R_\mathrm{csm}[\mathrm{cm}]$
\\
\hline
\verb|S_M01|&0.1&spherical&$5\times10^{15}$\\
\verb|S_M1|&1.0&spherical&$5\times10^{15}$\\
\verb|S_M10|&10.0&spherical&$5\times10^{15}$\\
\verb|D10_M01|&0.1&10&$5\times10^{15}$\\
\verb|D10_M1|&1.0&10&$5\times10^{15}$\\
\verb|D10_M10|&10.0&10&$5\times10^{15}$\\
\verb|D20_M01|&0.1&20&$5\times10^{15}$\\
\verb|D20_M1|&1.0&20&$5\times10^{15}$\\
\verb|D20_M10|&10.0&20&$5\times10^{15}$\\
\hline\hline
\end{tabular}
\label{table:models}
\end{center}
\end{table}

\subsection{1D spherical models}\label{sec:1d_spherical_models}
We first carry out 1D simulations with spherical symmetry. 
We employ the spherical SN ejecta (Equation \ref{eq:rho_sn}) and the spherical CSM (Equation \ref{eq:rho_csm}) with the CSM mass of $M_\mathrm{csm}=0.1$, $1.0$, and $10\ M_\odot$.

\subsubsection{Dynamical evolution}

\begin{figure}
\begin{center}
\includegraphics[scale=0.5]{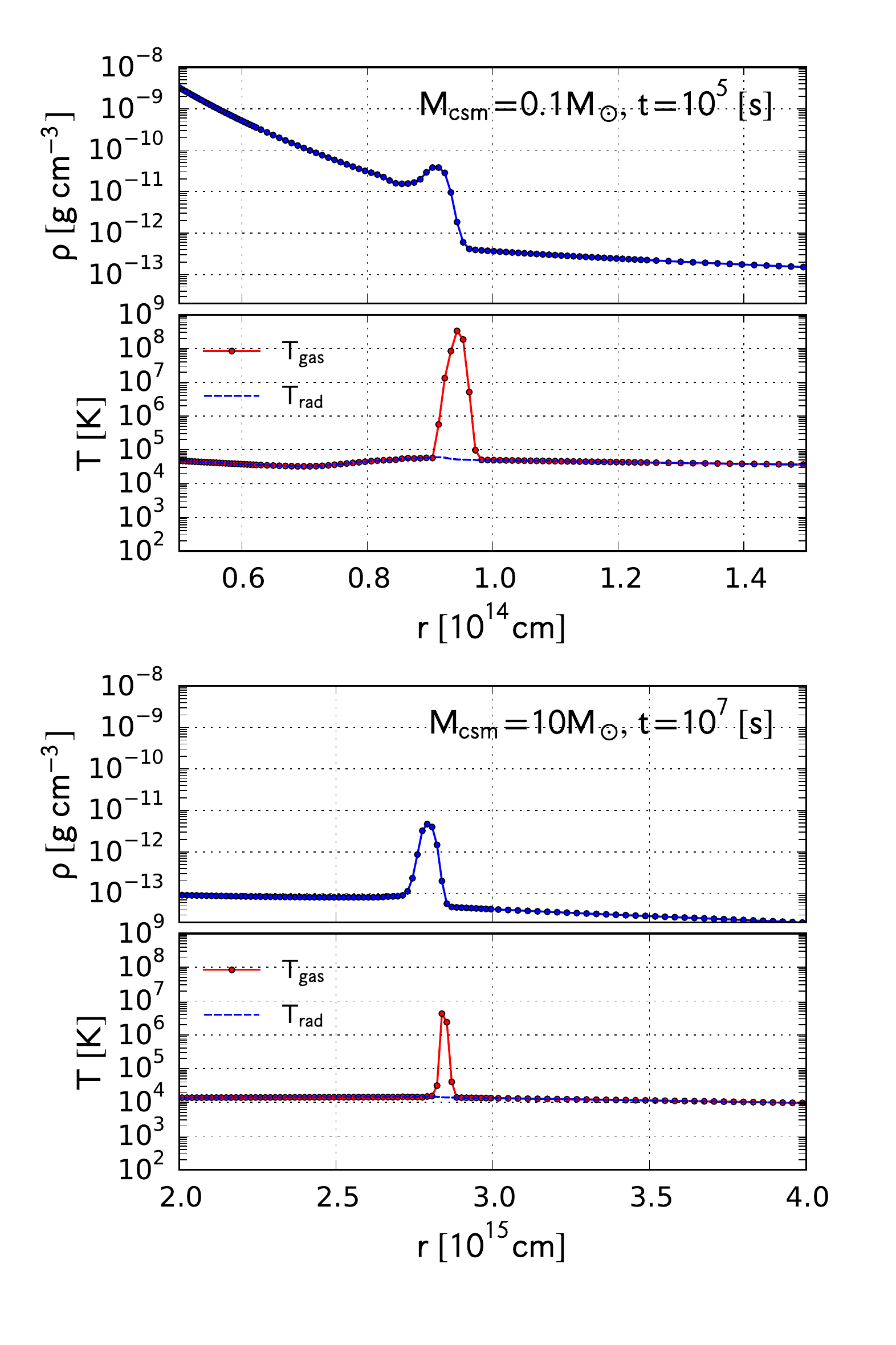}
\caption{Up-close views of the shock front. 
The upper and lower parts of the two panels correspond to the density and temperature profiles of the models with $M_\mathrm{csm}=0.1M_\odot$ at $t=10^5$ s and $M_\mathrm{csm}=10M_\odot$ at $t=10^7$ s, which are shown in Figures \ref{fig:radial_1d_m01} and \ref{fig:radial_1d_m10}, respectively. 
The centers of numerical cells are represented by circles in each panel. 
}
\label{fig:spike}
\end{center}
\end{figure}

In Figures \ref{fig:radial_1d_m01} and \ref{fig:radial_1d_m10}, we plot the radial profiles of several hydrodynamic variables obtained by the simulations with $M_\mathrm{csm}=0.1$ and $10\ M_\odot$, respectively. 
The simulations qualitatively reproduce early studies on SN ejecta and the associated blast wave evolution in a dense CSM \citep[e.g.,][]{2012ApJ...757..178G,2013MNRAS.428.1020M}. 
The shock front driven by the expanding ejecta sweeps the surrounding medium. 
Initially, the shock is almost adiabatic because of the short mean free path of photons in the dense medium. 
In this stage, the time required for the energy exchange between gas and radiation is much shorter than the dynamical timescale, resulting in the almost identical gas and radiation temperature distributions. 
At later epochs, however, the pre-shock gas density and the corresponding optical depth gradually decrease. 
This allows the post-shock radiation diffuse into the pre-shock region, to create the so-called ``precursor'', where the pre-shock gas is heated and accelerated by radiation. 
The precursor is followed by a layer with a very high gas temperature (known as Zel'dovich spike; see, e.g., \citealt{1967pswh.book.....Z}). 
This layer is immediately behind the shock front, at which the shock kinetic energy is directly converted into the internal energy of the post-shock gas. 
The internal energy of the post-shock gas is gradually shared by gas and radiation around the shock front, achieving gas-radiation equilibrium behind the shock front. 
When the precursor breaks out from the photosphere in the CSM, photons produced in the post-shock region can escape into the surrounding interstellar space. 
The hot layer behind the shock front has become wide because of the inefficient coupling between gas and radiation in the wind region, $R>R_\mathrm{csm}$. 

Figure \ref{fig:spike} shows the structure of the forward shock front. 
We plot the density and temperature profiles for the models with $M_\mathrm{csm}=0.1M_\odot$ at $t=10^5$ s and $M_\mathrm{csm}=10M_\odot$ at $t=10^7$ s. 
The high-temperature spike followed by cooled, piled-up gas layer is clearly recognized. 
The width of the high-temperature layer is roughly estimated as follows. 
The shock velocity at later epochs shown in Figures \ref{fig:snap_sph_m01} and \ref{fig:snap_sph_m10} is typically $v_\mathrm{shock}\sim 0.01c$. 
Therefore, the post-shock gas temperature can be as high as,
\begin{equation}
    T_\mathrm{g,shock}\simeq\frac{\mu m_\mathrm{u}v_\mathrm{shock}^2}{2k_\mathrm{B}}\simeq 3\times 10^9 \mathrm{K}. 
\end{equation}
Neglecting the radiation energy and relativistic effects, the cooling rate of the gas energy density is given by 
\begin{equation}
    \frac{d\bar{E}_\mathrm{g}}{dt}\simeq-c\bar{\rho}\bar{\kappa}_\mathrm{a}a_\mathrm{r}\bar{T}_\mathrm{g}^4,
\end{equation}
which leads to the following cooling timescale:
\begin{equation}
    t_\mathrm{cool}=\frac{\bar{E}_\mathrm{g}}{d\bar{E}_\mathrm{g}/dt}=
    \frac{k_\mathrm{B}}{(\gamma-1)\mu m_\mathrm{u}\bar{\kappa}_\mathrm{a}a_\mathrm{r}\bar{T}_\mathrm{g}^3}.
\end{equation}
Accordingly, the width $l_\mathrm{cool}$ of the relaxation layer is roughly estimated to be
\begin{eqnarray}
    l_\mathrm{cool}&\simeq&
    v_\mathrm{shock}t_\mathrm{cool}
    \nonumber\\&\simeq& 
    8\times 10^{12}
    \left(\frac{\rho}{10^{-11}\ \mathrm{g}\ \mathrm{cm}^{-3}}\right)^{-1}
    \left(\frac{v_\mathrm{shock}}{0.01c}\right)^2\ \mathrm{cm}.
\end{eqnarray}
Taking $\rho=10^{-11}$ g cm$^{-3}$ and $v_\mathrm{shock}=0.01c$ for example, this order of magnitude estimation gives $l_\mathrm{cool}\sim 10^{13}$ cm, which explains the width of the high-temperature spike in the upper panel of Figure \ref{fig:spike}. 
The spike in the lower panel is wider than that in the upper panel, probably reflecting its lower post-shock density. 
In reality, the density of the gas flow following the spike widely varies due to the piling-up of material, which makes it difficult to correctly estimate the width of the high-temperature layer. 
However, the above order of magnitude estimation seems to work well. 
As seen in Figure \ref{fig:spike}, the high-temperature spikes are covered by less than 10 numerical cells. 
Although we avoid a spike represented by only a single numerical cell, the shock structure is likely smeared out to some extent.

For the model with $M_\mathrm{csm}=0.1M_\odot$, the CSM mass is much smaller than the total ejecta mass of $M_\mathrm{ej}=10M_\odot$. 
Therefore, only a small fraction of the outer ejecta is swept by the reverse shock front. 
The maximum velocity of the ejecta before the shock emergence from the outer edge of the CSM ($t=10^6$ s) is $\simeq 6\times 10^3$ km s$^{-1}$. 
Then, after the emergence, the shock front accelerates again due to the steep density gradient around the outer edge of the CSM. 

For the model with $M_\mathrm{csm}=10M_\odot$, on the other hand, the total mass of the CSM is comparable to the ejecta. 
Thus, the ejecta significantly decelerate and dissipate a considerable fraction of the kinetic energy. 
As seen in Figure \ref{fig:radial_1d_m10}, the maximum velocity of the ejecta decreases down to $2\times 10^3$ km s$^{-1}$ at $t=10^7$ s. 
The forward shock front has not reached the outer edge of the CSM even at $t=10^7$ s, and thus still contributes to the bright thermal emission. 
In this model, a dense and geometrically thin shell is formed. 
The density of the shell is higher than that of the surrounding unshocked gas by about two orders of magnitude, which cannot be achieved by the adiabatic shock jump condition alone. 
This is the so-called cooling shell, in which the effective energy loss makes material piling up in the layer between the forward and reverse shock fronts.

\subsubsection{Light curves}
\begin{figure}
\begin{center}
\includegraphics[scale=0.55]{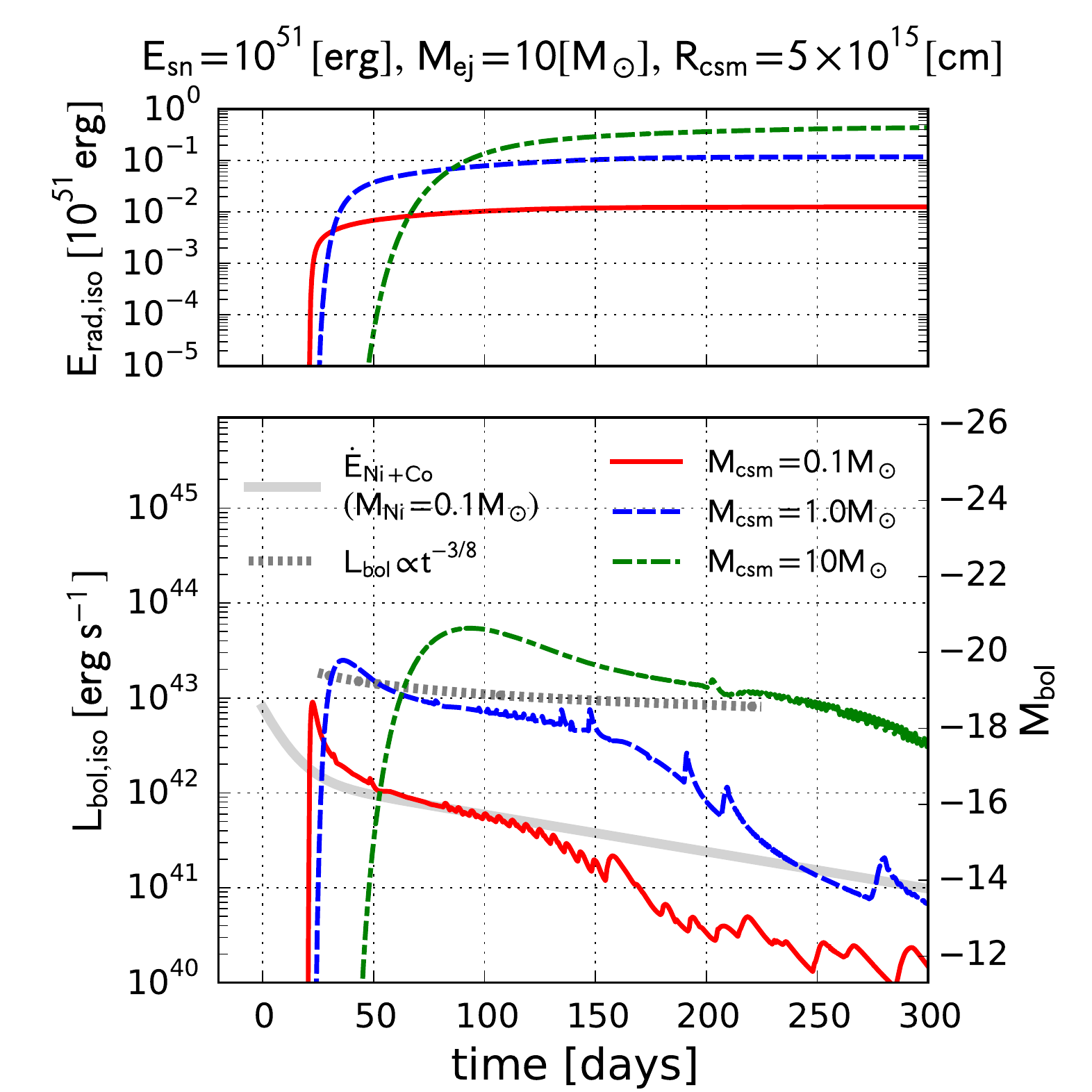}
\caption{Bolometric light curves of the 1D spherical models with $M_\mathrm{csm}=0.1$, $1.0$, and $10\ M_\odot$. 
In the upper panel, the cumulative light curves are also plotted. 
}
\label{fig:lc_1d}
\end{center}
\end{figure}

The bolometric light curves of the 1D spherical models are presented in Figure \ref{fig:lc_1d}. 
In the upper panel, we plot the cumulative radiated energy, Equation (\ref{eq:E_rad}). 
In the lower panel, the light curves are compared with each other and the radioactive energy deposition rate $\dot{E}_\mathrm{Ni+Co}$ with the nickel mass of $M_\mathrm{Ni}=0.1M_\odot$ \citep{1994ApJS...92..527N}. 
We note that some spikes in the light curves are artificially produced due to numerical treatments, especially the non-uniform AMR grid structure (this is alleviated in 2D simulations, see below).
Nevertheless, the general trends of the light curves around the peak luminosity are nicely captured.

The CSM mass governs the peak and the characteristic timescale of the light curve evolution.
For the lower CSM mass model, $M_\mathrm{csm}=0.1M_\odot$, the contribution of the CSM interaction is very limited. 
It certainly contributes to the light curve within $\sim10$ days around the peak, at which the forward shock is still in the CSM. 
At later epochs, the luminosity continuously decays down to values below the radioactive energy deposition rate. 
The contribution of the CSM interaction becomes more significant for increasing CSM mass. 
The models with larger CSM masses are more capable of dissipating the kinetic energy of the SN ejecta, but it takes longer times.  
The peak luminosity and the total radiated energy plotted in Figure \ref{fig:lc_1d} indeed increase for increasing CSM mass. 
At the same time, the elevated CSM densities make the rise and decay timescales longer. 
At the beginning, the CSM interaction occurs at a deep layer of the CSM, which is initially hidden from the observer. 
The higher CSM density and thus the larger optical depth lead to a longer diffusion time. 
Therefore, it  takes longer times for photons produced in the interaction layer to emerge from the outer edge of the CSM. 
As a result, for the model with the highest CSM mass, bright emission with $L_\mathrm{bol,iso}\ga10^{43}$ erg s$^{-1}$ lasts for more than 100 days. 
A more quantitative discussion on the rise and decline timescales is found in Section \ref{sec:rise_time}. 
The total radiated energy reaches terminal values of $E_\mathrm{rad,iso}=1.3\times10^{49}$, $1.2\times 10^{50}$, and $4.4\times 10^{50}$ erg for the models with $M_\mathrm{csm}=0.1$, $1.0$, and $10M_\odot$, corresponding to the conversion efficiencies of $1.3$, $12$, and $44\%$.

While the outer part of the SN ejecta ($\rho\propto r^{-m}$) is interacting with a CSM with a power-law radial density profile with an index $-s$, $\rho\propto R^{-s}$, the kinetic energy dissipation rate is proportional to $t^{(6s-15+2m-ms)/(m-s)}$ \citep{2013MNRAS.435.1520M}. 
For the adopted parameter set, the exponent is found to be $-3/8$. 
When the dissipated kinetic energy is assumed to be converted to radiation at a constant rate, the light curve of the interaction-powered emission roughly follows this time dependence at timescales longer than the photon diffusion time in the CSM. 
Therefore, the bolometric luminosity behaves as $L_\mathrm{bol}\propto t^{-3/8}$. 
In Figure \ref{fig:lc_1d}, we compare this power-law time dependence with the 1D spherical model with $M_\mathrm{csm}=1.0M_\odot$. 
After the bump around the maximum light, at which the timescale of the light curve evolution is determined by the photon diffusion timescale, the light curve flattens and decays at a similar rate to the power-law function. 
The numerical light curve is declining more rapidly than the power-law time dependence. 
This may be caused by a time dependent conversion efficiency. 
Recently, \cite{2019arXiv190705166T} developed a light curve model for interacting SNe by including the effect of a time-dependent conversion efficiency. 
They claim that the time-dependent conversion efficiency can make light curves decline more rapidly. 
They also found that the forward and reverse shock components behave differently due to the different post-shock densities, resulting in a double power-law light curve. 
In the earlier part, where the forward shock dominates, the luminosity decreases more rapidly than $t^{-3/8}$ due to the time-dependent conversion efficiency.

\subsection{2D Spherical Models}\label{sec:spherical_csm}
We have carried out 2D cylindrical simulations with spherical CSMs. 
The CSM mass is set to $M_\mathrm{csm}=0.1$, $1.0$, and $10\ M_\odot$, corresponding to the spherical 1D models in Section \ref{sec:1d_spherical_models}. 

\subsubsection{Dynamical evolution}\label{sec:spherical_csm_dynamical}
\begin{figure*}
\begin{center}
\includegraphics[scale=0.2]{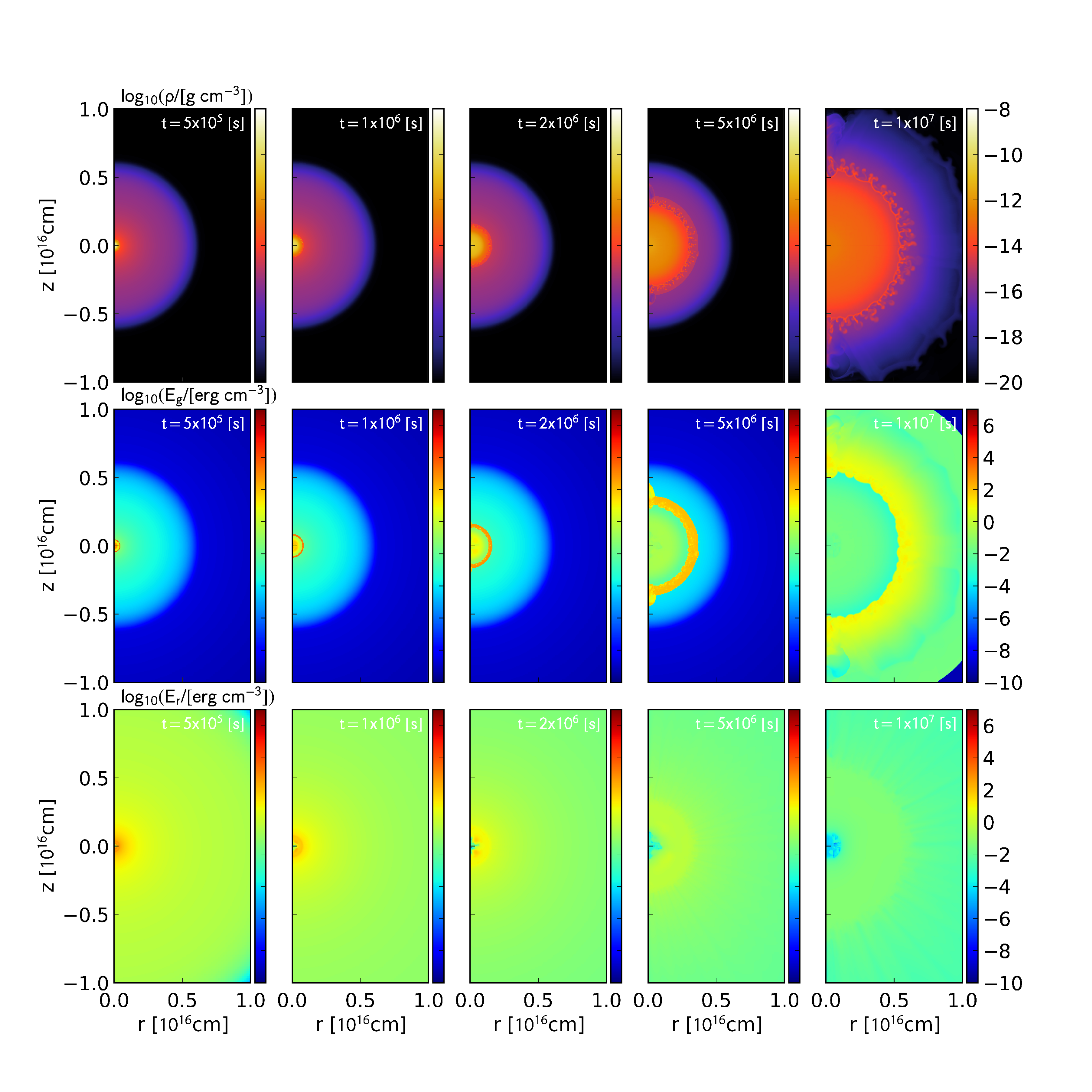}
\cprotect\caption{Results of the 2D spherical CSM model with $M_\mathrm{csm}=0.1M_\odot$ (model \verb|S_M01|). 
Spatial distributions of the density (upper panels), the gas energy density (middle panels), and the radiation energy density (lower panels) are presented. 
The columns represent the spatial distributions at $t=5\times 10^5$, $10^6$, $2\times 10^6$, $5\times 10^5$, and $10^7$ s, from left to right. 
}
\label{fig:snap_sph_m01}
\end{center}
\end{figure*}
\begin{figure*}
\begin{center}
\includegraphics[scale=0.2]{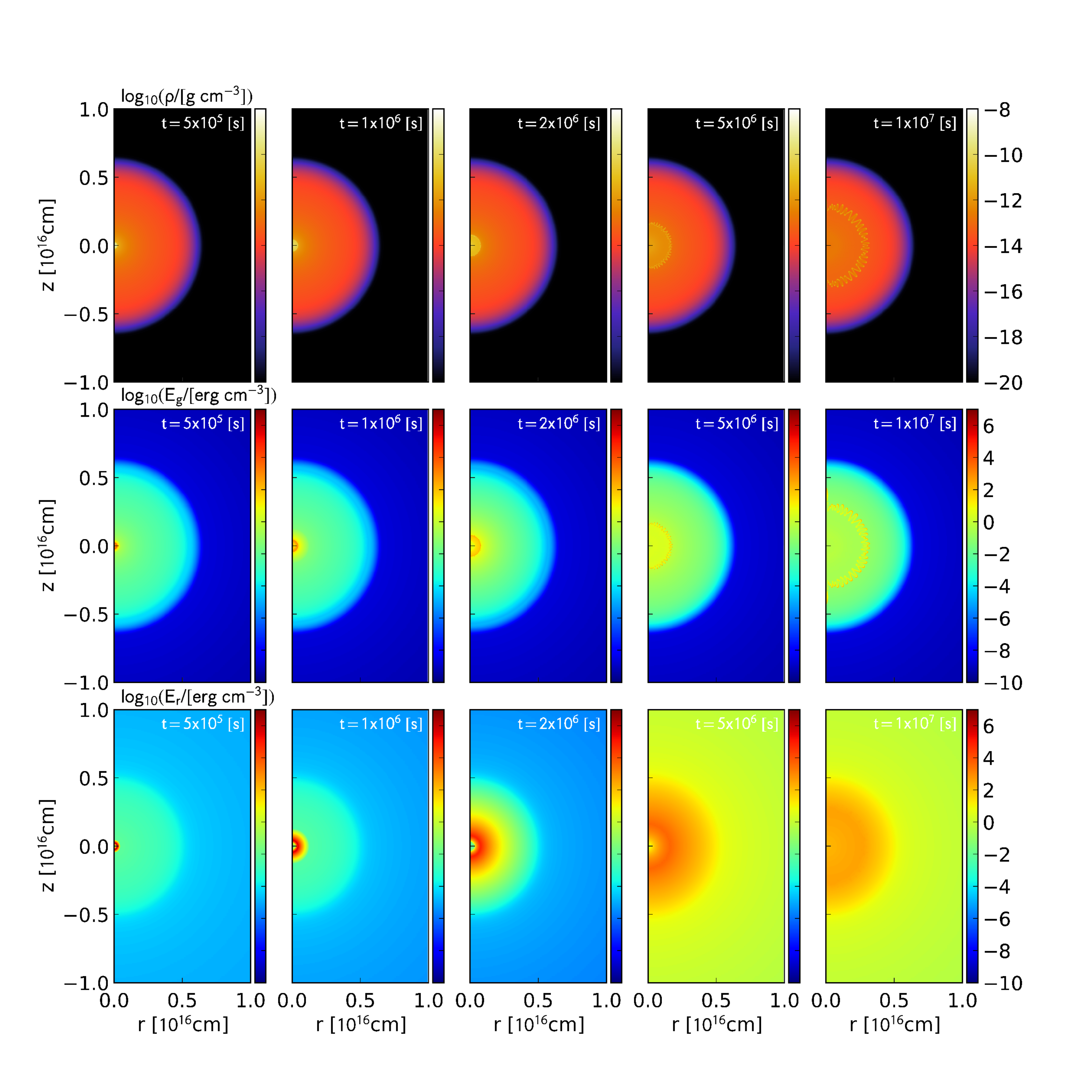}
\cprotect\caption{Same as Figure \ref{fig:snap_sph_m01}, bur for the 2D spherical CSM model with $M_\mathrm{csm}=10M_\odot$ (model \verb|S_M10|). }
\label{fig:snap_sph_m10}
\end{center}
\end{figure*}

Figures \ref{fig:snap_sph_m01} and \ref{fig:snap_sph_m10} show the temporal evolutions of the SN ejecta interacting with the CSM with $M_\mathrm{csm}=0.1$ and $10M_\odot$ (models \verb|S_M01| and \verb|S_M10|). 
Their dynamical evolutions are similar to the 1D models except for the ejecta-CSM interface. 
In 2D simulations, the contact surface is subject to hydrodynamic instabilities. 
As we have noted in Section \ref{sec:1d_spherical_models}, for $M_\mathrm{csm}=0.1M_\odot$, the radiative precursor develops even at several $10^4$ s after the explosion. 
This can also be seen in the spatial distributions of the radiation energy density (bottom panels of Figure \ref{fig:snap_sph_m01}). 
The radiation front propagates well ahead of the shock front that is still deeply embedded in the CSM, resulting in the pre-shocked region with a high radiation energy density $E_\mathrm{r}> 1$ erg cm$^{-3}$. 
On the other hand, for the model with the highest CSM mass (the bottom panels of Figure \ref{fig:snap_sph_m10}), the region with high radiation energy density is well confined in the massive CSM at early epochs (up to several $10^6$ s). 
The radiation front then breaks out from the outer edge of the CSM and fill the surrounding space with radiation. 
We note the shock front is still in the CSM even at $t=10^7$ s. 

One of the important hydrodynamic features in these 2D simulations is the Rayleigh-Taylor instability. 
The forward shock front accumulates increasing amounts of the surrounding medium and thus decelerates as it propagates in the CSM. 
Because of the deceleration, the shocked SN ejecta feel the inertial force toward the direction of the shock propagation, which tries to replace dense gas in the shocked ejecta with relatively dilute gas in the shocked CSM. 
As a result, Rayleigh-Taylor fingers appear at the ejecta-CSM interface and travel into the shocked CSM. 
The shortest unstable wavelength of the Rayleigh-Taylor instability is determined by several physical conditions including diffusion length \citep[e.g.,][]{1978ApJ...219..994C}. 
In these simulations, however, the instability develops from numerical errors, mainly caused by AMR grid structures and the physically determined shortest wavelength is not resolved even with AMR, especially at early epochs. 

\begin{figure}
\begin{center}
\includegraphics[scale=0.1]{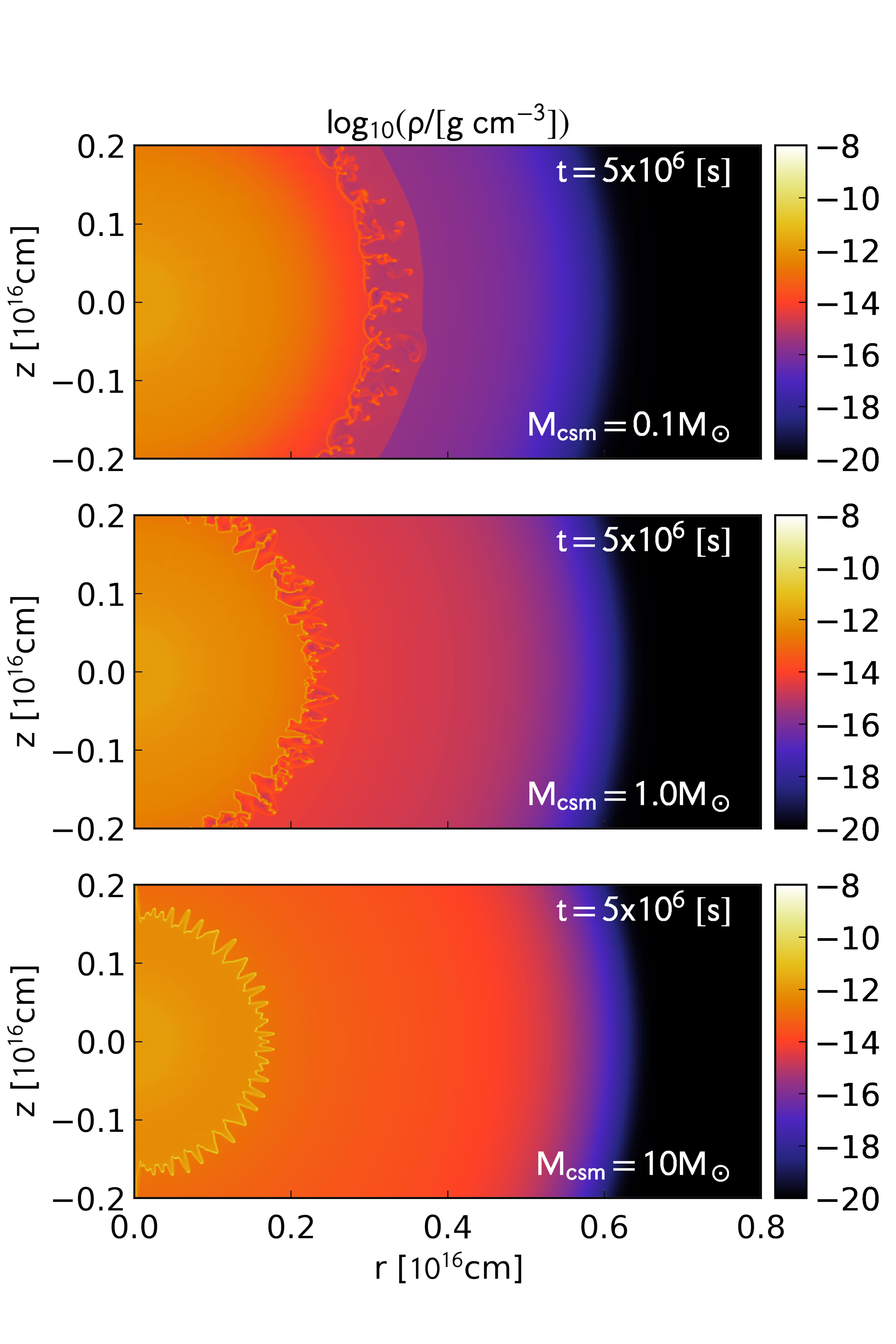}
\caption{Comparison of the shock structures in the 2D spherical CSM models. 
The density distributions of the spherical CSM models at $5\times 10^6$ s with $M=0.1$ (top), $1.0$ (middle), and $10\ M_\odot$ (bottom) are presented. 
}
\label{fig:shock}
\end{center}
\end{figure}

Another instability potentially playing a role in this setting is the Vishniac instability (also known as the non-linear thin shell instability). 
\cite{1983ApJ...274..152V} and \cite{1987ApJ...313..820R} analytically considered the stability of a blast wave traveling in a uniform medium and found that it is overstable when the effective adiabatic index is close to unity ($\gamma_\mathrm{eff}<1.2$). 
In the Vishniac instability, the wave front or the shell oscillates with an amplitude increasing with time in a power-law fashion. 
In a radiative shock, a part of the dissipated shock kinetic energy is lost as escaping radiation, effectively reducing the adiabatic index \citep{1986ApJ...304..154B,1993ARA&A..31..373D}. 
Later numerical studies confirm that this kind of instability certainly occurs in radiative shocks in various astrophysical phenomena, including SN remnants in the radiative phase \citep[e.g.,][]{1998ApJ...500..342B,2012ApJ...759...78M,2016MNRAS.459.2188B,2018A&A...617A.133M}.

These two instabilities both make the shock front deviate from its initial spherical shape. 
While the Rayleigh-Taylor instability is a purely hydrodynamic instability that can develop in any stages of the dynamical evolution, the Vishniac instability plays a role only when the radiative loss in the post-shock region is significant and the effective adiabatic index of the post-shock gas is below a threshold value. 
When the optical depth from the outer edge of the CSM to the shock front is sufficiently large, the total pressure is dominated by the radiative one and thus the adiabatic index of the post-shock gas becomes close to that of photon gas, i.e., $\gamma=4/3$. 
As the forward shock approaches the outer edge of the CSM, the radiative loss becomes increasingly significant. 
Then, the mixture of the gas and radiation behaves as a gas with an adiabatic index close to unity. 
This is when the Vishniac instability possibly develops. 
For lower CSM densities, however, insignificant radiative loss prevents the Vishniac instability from developing even at later epochs. 
In our simulations, the free-free process is the only radiative process for direct gas cooling. 
Therefore, for lower CSM densities with insignificant radiative cooling, gas should behave as an ideal gas with an adiabatic index of $\gamma=5/3$ as assumed and the Vishniac instability is not effective within the timescale covered by the simulations.

In Figure \ref{fig:shock}, we compare the shock structure at $t=5\times 10^{6}$ s in the three 2D spherical CSM models. 
In the spatial distribution shown in the upper panel of Figure \ref{fig:shock}, the ejecta-CSM interface indeed exhibits filamentary structure, which is likely created as a result of the Rayleigh-Tayler instability. 
In contrast to the models with higher CSM densities, the forward shock front and the contact discontinuity are well separated from each other. 
This suggests that the post-shock gas almost behaves as an ideal gas without significant radiative loss. 
Under these circumstances,  the motion of gas around the ejeca-CSM interface can be turbulent, leading to additional dissipation of the kinetic energy due to collisions of filaments.  
In most parts, however, the filaments and the perturbed shell are confined in the narrow shocked layer between the forward and reverse shock fronts. 
Even at several $10^7$ s after the explosion, the development of the hydrodynamic instability is not strong enough to modify the global shape of the ejecta. 
We have also checked that the radial profiles of hydrodynamic variable are similar to those in 1D spherical models. 
Thus, the effect of the additional energy dissipation would be quite limited as long as these particular simulations are concerned. 
As we shall see below, the light curves of 2D spherical models are not significantly different from the corresponding 1D spherical models. 
However, one caveat is that our simulations assume no initial density and velocity perturbations and the instabilities develop from small numerical errors. 
We cannot exclude a possibility that large density and velocity perturbations are present in pre-explosion stars and/or CSMs and enhance the development of hydrodynamics instabilities after the explosion. 
In such cases, its influence on the energy dissipation and light curves is expected to be more significant.

In the bottom panel of Figure \ref{fig:shock}, the shocked gas in model \verb|S_M10| seems to suffer from significant radiative loss. 
Unlike the models with the smaller CSM masses, the forward shock front and the contact discontinuity are not well separated. 
The density distribution shows geometrically thin shell as in 1D spherical models, although its shape is far from spherical. 
The creation of a cooling shell suggests that radiative cooling is very effective in this particular model. 
The non-spherical cooling shell is probably due to the development of the Vishniac instability. 

\subsubsection{Light curves}
\begin{figure}
\begin{center}
\includegraphics[scale=0.55]{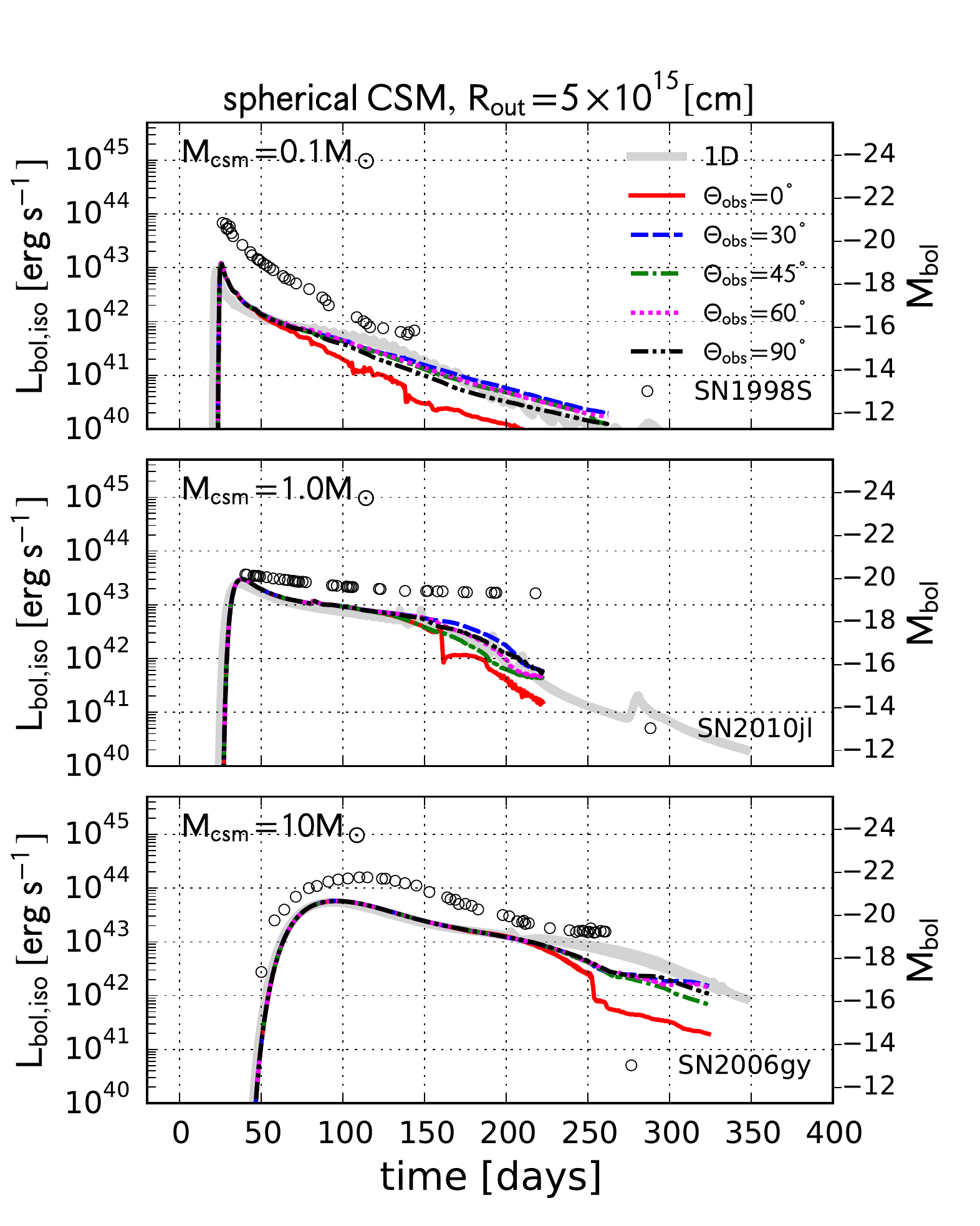}
\caption{Isotropic equivalent bolometric light curves of the 2D spherical CSM models with $M_\mathrm{csm}=0.1$, $1.0$, and $10\ M_\odot$, from top to bottom. 
The light curves with viewing angles of $0^\circ$, $30^\circ$, $45^\circ$, $60^\circ$, and $90^\circ$ are plotted and compared with the corresponding 1D spherical model in each panel. 
We also plot light curves of some type IIn SNe whose peak bolometric luminosities or decaying trends are similar to the theoretical light curves, SN 1998S \citep[top;][]{2000MNRAS.318.1093F}, 2010jl \citep[middle;][]{2012AJ....144..131Z}, and 2006gy (bottom). 
For SN 2006gy, we show the $R$-band light curve with no bolometric correction \citep[][]{2010ApJ...709..856S}. 
The Galactic and host galaxy extinctions are assumed to be $A_{R,\mathrm{mw}}=0.43$ and $A_{R,\mathrm{host}}=1.25$. 
}
\label{fig:lc_sph}
\end{center}
\end{figure}

In Figure \ref{fig:lc_sph}, we plot isotropic equivalent bolometric light curves with different viewing angles ($\Theta_\mathrm{obs}=0^\circ$, $30^\circ$, $45^\circ$, $60^\circ$, and $90^\circ$) and compare them with those of 1D spherical simulations. 
For most parts, the light curves with different viewing angles are similar to each other and the 1D spherical counterparts at early epochs, although small differences are recognized at later epochs. 
At around the maximum light, photons produced in the ejecta--CSM interface predominantly contribute to the emission. 
Such photons would have experienced multiple absorption and scattering episodes, which make the radiation field nearly isotropic. 
This explains why the light curves are similar to each other at early epochs. 
In contrast, at later epochs, photons from the non-spherical interaction layer escape into the surrounding space without significant absorption and scattering.  
Then, light curves with different viewing angles start deviating from each other. 
Nevertheless, the difference between the light curves shown in each panel of Figure \ref{fig:lc_sph} is not significant. 

First, we note that the light curves with $\Theta_\mathrm{obs}=0^\circ$ are exceptionally different from those with different viewing angles. 
Caveat is that this feature is probably because of an artificial effect associated with the assumed axi-symmetry. 
We impose a reflecting boundary condition at the symmetry axis. 
Therefore, incoming gas motion toward the reflecting boundary encounters the opposite gas motion with the same amplitude, enhancing existing perturbations and numerical errors. 
We indeed recognize artificial structure around the symmetry axis in the density distributions shown in Figures \ref{fig:snap_sph_m01} and \ref{fig:snap_sph_m10}. 
Thus, this deviation would probably be resolved in three-dimensional simulations.

Next, we mention the systematic difference in light curves with other viewing angles (e.g. the light curves viewing from 30 degrees is always larger than those of 45 degrees). 
In 2D cylindrical geometry, the development of the Raleigh-Taylor instability along a particular radial direction systematically depends on the inclination angle, which causes the differences in the light curves, because perturbations having developed in the simulation are not actually clumps but rings. 
This makes the development of perturbations in different orientations systematically different. 
Since the differences in late-time light curves are probably due to different ways of the energy dissipation at the interaction layer at each direction, the late-time light curves are certainly influenced by this systematic effect. 
These late-time differences would behave correctly in 3D simulations. 
However, the deviations of the late-time light curves from each other is within a factor of a few except for $\Theta_\mathrm{obs}=0^\circ$. 
The peak bolometric luminosities of 2D spherical models are larger than the corresponding 1D spherical models are at most $37$, $22$, and $7\%$ for models with $M_\mathrm{csm}=0.1$, $1.0$, and $10M_\odot$, respectively. 
This indicates that the effect of the non-spherical interaction layer due to the Rayleigh-Taylor instability is limited. 
At least, the instability does not change the luminosity by an order of magnitude. 
As we have noted in Section \ref{sec:spherical_csm_dynamical}, however, pre-existing large density and velocity perturbations may enhance the development of the instability to produce global non-spherical structure. 
In such cases, the luminosity could vary more widely because of the additional energy dissipation by ejecta-clump interaction.

In Figure \ref{fig:lc_sph}, we also plot light curves of some type IIn SNe, 1998S, 2010jl, and 2006gy, in the literature. 
Although it is not our purpose here to precisely reproduce light curves of particular interacting SNe by adjusting some model parameters, this comparison clearly demonstrates that there are some type IIn SNe with their light curve shapes similar to the simulations results.  

In summary, deviations from 1D spherical models are not significant.
Therefore, we can conclude that effects of multi-dimensional hydrodynamic instabilities on light curves are negligible at least for spherical SN ejecta interacting with a spherical CSM. 

\subsection{Disk CSM}\label{sec:disk_csm}

More drastic multi-dimensional effects are expected in the presence of aspherical CSMs. 
We perform simulations assuming disk-like CSMs with different CSM masses and opening angles.

\subsubsection{Dynamical evolution}
\begin{figure*}
\begin{center}
\includegraphics[scale=0.20]{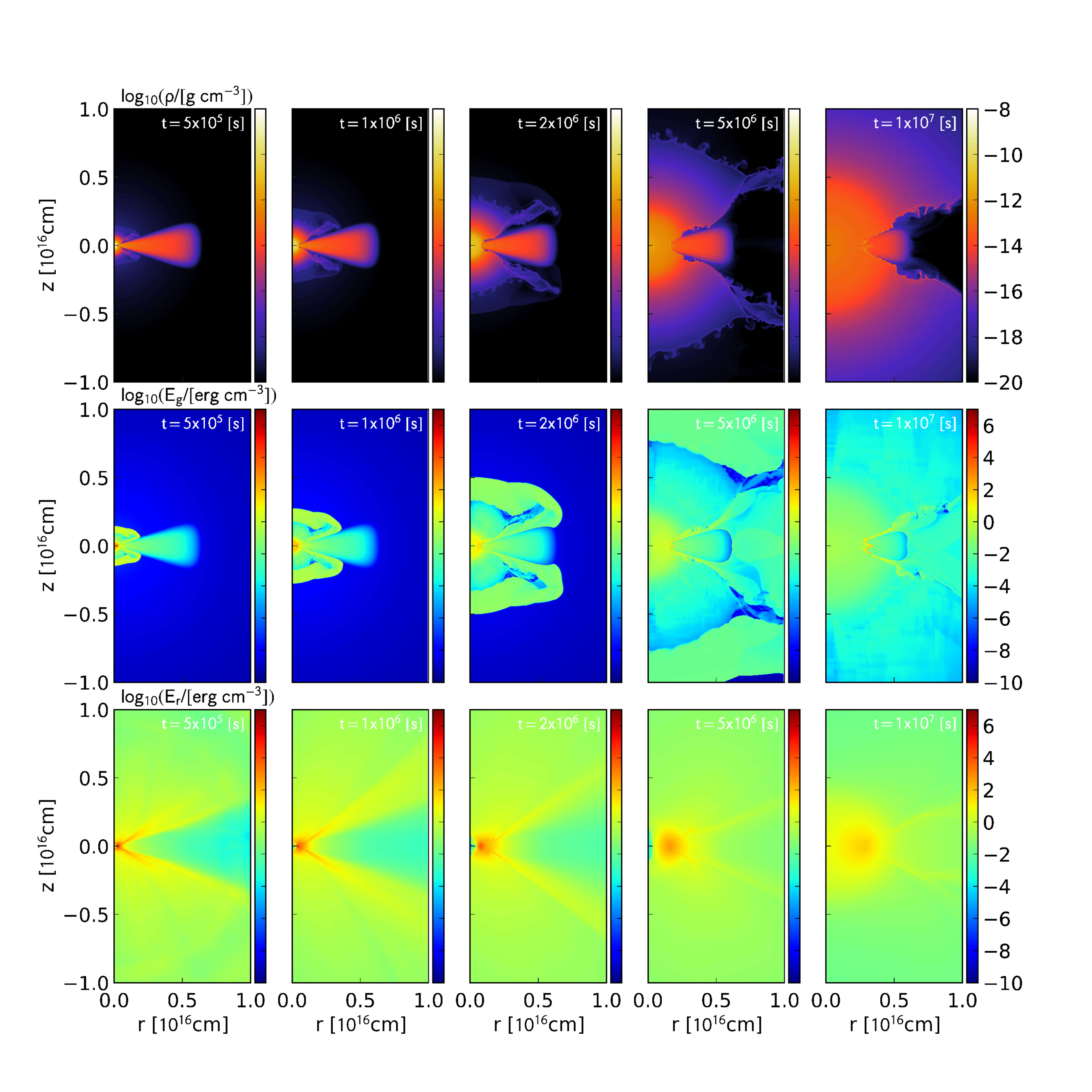}
\cprotect\caption{Results of model \verb|D10_M1|. 
Spatial distributions of the density (upper panels), the gas energy density (middle panel), and the radiation energy density (lower panel) are plotted. 
The columns represent the spatial distributions at $t=5\times 10^5$, $10^6$, $2\times10^6$, $5\times 10^6$, $10^7$ s from left to right. }
\label{fig:snap_disk1}
\end{center}
\end{figure*}
\begin{figure*}
\begin{center}
\includegraphics[scale=0.20]{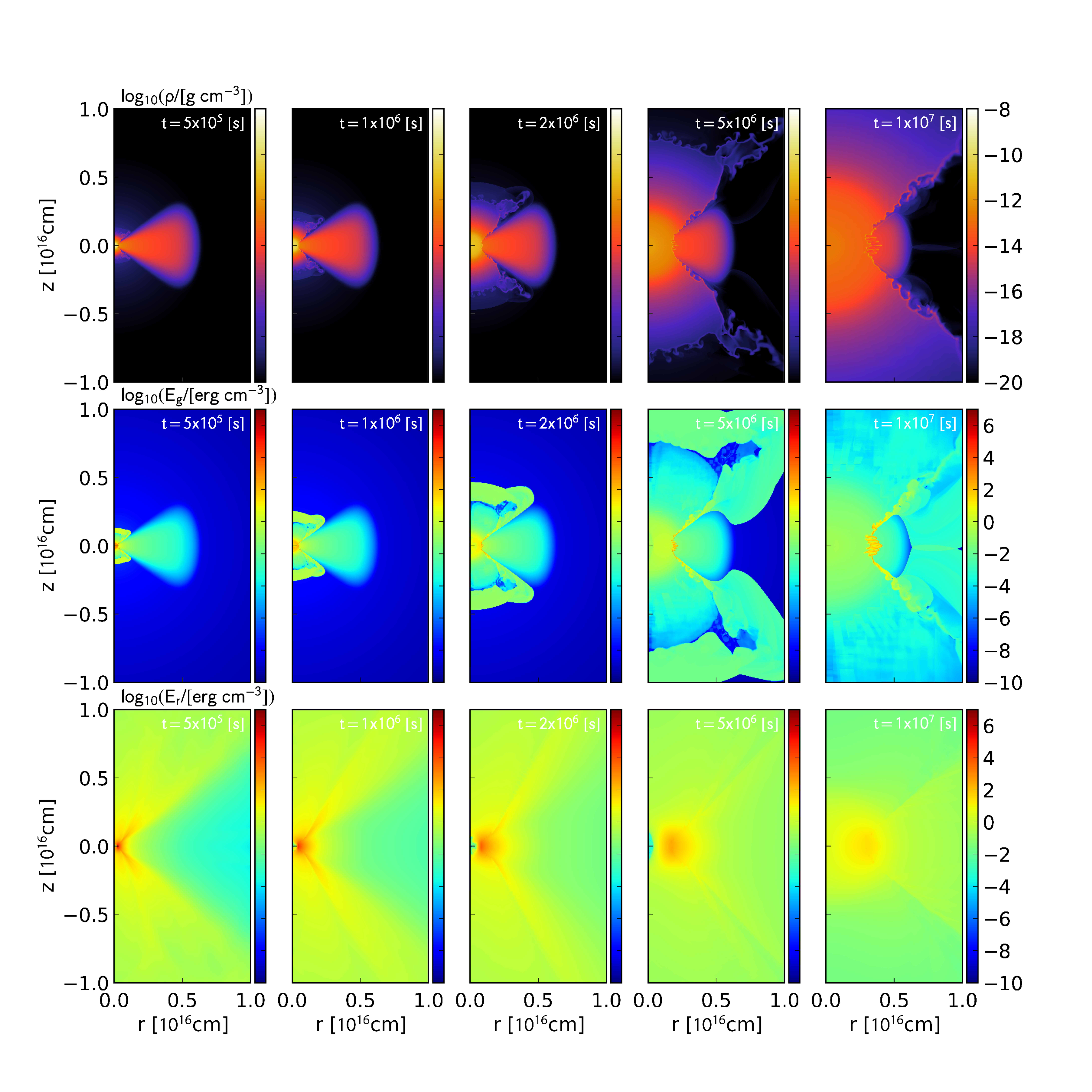}
\cprotect\caption{Same as Figure \ref{fig:snap_disk1}, but for model \verb|D20_M1| }
\label{fig:snap_disk2}
\end{center}
\end{figure*}

Figures \ref{fig:snap_disk1} and \ref{fig:snap_disk2} present the dynamical evolution of the ejecta-CSM interaction in models \verb|D10_M1| and \verb|D20_M1|. 
The ejecta start interacting with the surrounding media immediately after the beginning of the simulations. 
In the presence of a CSM disk, a part of the ejecta traveling around the equator decelerates efficiently, resulting in highly aspherical ejecta structure (top panels of Figures \ref{fig:snap_disk1} and \ref{fig:snap_disk2}). 
The polar part of the ejecta is not covered by the CSM and thus the ejecta can expand almost freely. 
As a result, the ejecta is squeezed in the presence of the CSM disk and a ring-like region with no radially expanding ejecta appears behind the CSM disk (hereafter, referred to as the void region). 

One of the important consequences of this ejecta--disk interaction is that the kinetic energy dissipation is most efficiently happening in the equator, where the ejecta is adjacent to the CSM disk. 
This is clearly seen in the spatial distributions of the radiation energy density, particularly in the bottom panels of Figures \ref{fig:snap_disk1} and \ref{fig:snap_disk2}. 
The region with the highest radiation energy density is found at an off-center location on the equatorial plane.  

Another important issue for the emission from SN ejecta is the impact of the viewing angle.
As we will see below, light curves of an SN interacting with a CSM disk are significantly affected by whether they are seen directly or through the CSM. 
Therefore, what fraction of the ejecta is affected by the CSM disk plays a critical role in determining the expected emission powered by the ejecta--disk interaction. 
From the density distributions in Figures \ref{fig:snap_disk1} and \ref{fig:snap_disk2}, there is a general trend that SN ejecta colliding with CSM disks with larger opening angles $\vartheta_\mathrm{disk}$ more significantly suffer from the squeezing effect. 
As a result, the solid angle corresponding to an almost freely expanding part of the SN ejecta appears to be a decreasing function of the disk half opening angle. 
It is expected that the corresponding solid angle is roughly given by $4\pi[1-F_q(\vartheta_\mathrm{disk})]$, where $F_q(\vartheta_\mathrm{disk})$ is the covering fraction of the CSM disk (Equation \ref{eq:F_q}). 

To be more precise, however, the boundary between the SN ejecta and the void region is determined by a more complicated hydrodynamics process. 
As seen in the top panels, the density distributions, of Figure \ref{fig:snap_disk2}, the inner part of the CSM disk is swept by the forward shock immediately after it starts interacting with the SN ejecta.
As a result, the inner disk is compressed and the associated pressure increase should also affect the ejecta that should have avoided the effect of the CSM disk in ballistic collision case. 
In this way, a part of the kinetic energy of the ejecta dissipated around the equator is redistributed into ejecta components close to the void region. 
In the density distributions shown in Figures \ref{fig:snap_disk1} and \ref{fig:snap_disk2}, the ejecta component traveling along the boundary between the ejecta and the void region precede those along the symmetry axis. 
This can be understood as a consequence of the complex hydrodynamic interaction between SN ejecta and a CSM disk, which cannot be achieved by considering a ballistic collision alone. 
The opening angle of the void region indeed seems to be slightly larger than the assumed half opening angle of the CSM disk. 

\begin{figure}
\begin{center}
\includegraphics[scale=0.42]{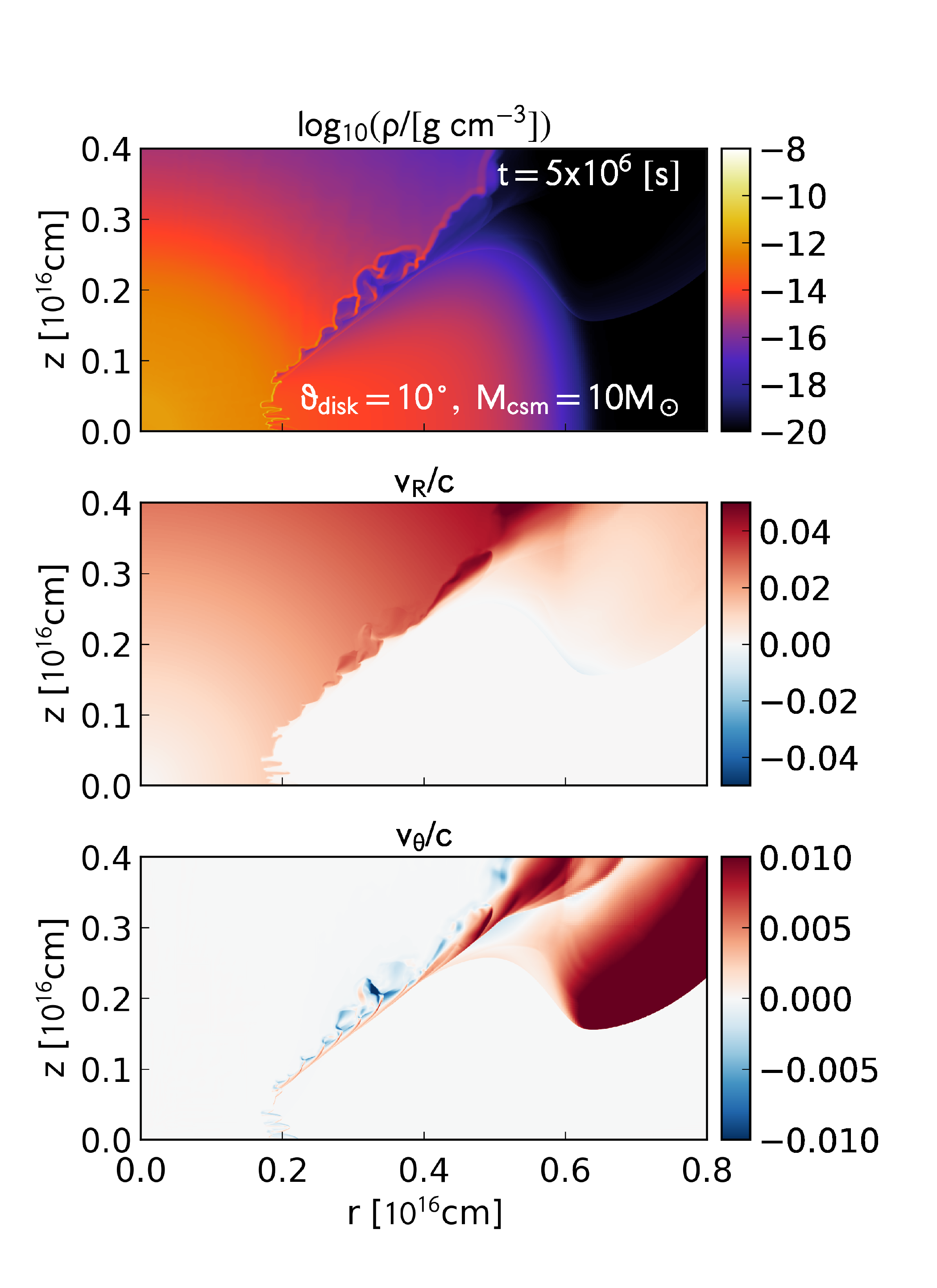}
\cprotect\caption{Structure of the ejecta-disk interface for model \verb|D20_M1| at $5\times 10^6$ s. 
The spatial distributions of the density (top), the radial velocity $v_R$ (middle), and the angular velocity $v_\theta$ (bottom) are plotted. 
}
\label{fig:shear_layer}
\end{center}
\end{figure}

In addition, the radially traveling ejecta adjacent to the disk surface could suffer from Kelvin-Helmholtz instability, leading to further dissipation of the kinetic energy. 
Figure \ref{fig:shear_layer} shows the density and velocity structures of the interface between the ejecta and the disk for model \verb|D20_M1|. 
The radial and angular components of the velocity at coordinates $(r,z)$ are defined as,
\begin{equation}
    v_R=v_r\frac{r}{R}+v_z\frac{z}{R},
\end{equation}
and
\begin{equation}
    v_\theta=v_r\frac{r}{R}-v_z\frac{z}{R}.
\end{equation}
The density distribution around the ejecta-disk interface ($2\times 10^{15}\mathrm{cm}\leq r \leq 4\times 10^{15}\mathrm{cm}$) exhibits nearly evenly-spaced perturbations with non-zero angular velocities. 
Since the ejecta-disk interface is a shear layer, where the radially expanding gas is adjacent to the CSM nearly at rest (the middle panel of Figure \ref{fig:shear_layer}), this perturbed interface is likely produced by Kelvin-Helmholtz instability. 
The development of the instability makes the ejecta around the interface clumpy, which would affect the emission traveling around the ejecta-disk interface. 
Although it seems to be the case in this particular simulation, it is not clear whether the instability always develops. 
The growth of the Kelvin-Helmholtz instability is sensitive to the density and velocity gradients at the shear layer, namely the vertical structure of the CSM disk. 
As long as the pre-supernova mass-loss process responsible for creating confined CSM is unclear, the role played by shear layers is also uncertain. 

As we have discussed, some hydrodynamic instabilities, the Rayleigh-Taylor, Vishniac, and possibly Kelvin-Helmholz instabilities, produce
small-scale structures in the ejecta-CSM interface. 
These instabilities develop even from numerical errors and thus destroys the equatorial symmetry. 
As seen in Figures \ref{fig:snap_disk1} and \ref{fig:snap_disk2}, the spatial distributions of physical variables in the upper and lower hemispheres certainly exhibit small equatorial asymmetries, although their global structures are similar. 
This leads to slight differences in light curves seen from upper and lower hemispheres as we shall see below.

We also note that the treatment of the upper and lower edge of the CSM disk is also important for the opening angle of the void region. 
In our simulations, we assume the CSM density structure in Equation (\ref{eq:rho_disk}) with $q=4$. 
However, the realistic density structure of the envelope of a CSM disk would likely depend on its formation process and therefore is highly uncertain.

\subsubsection{Light curves}

\begin{figure*}
\begin{center}
\includegraphics[scale=0.46]{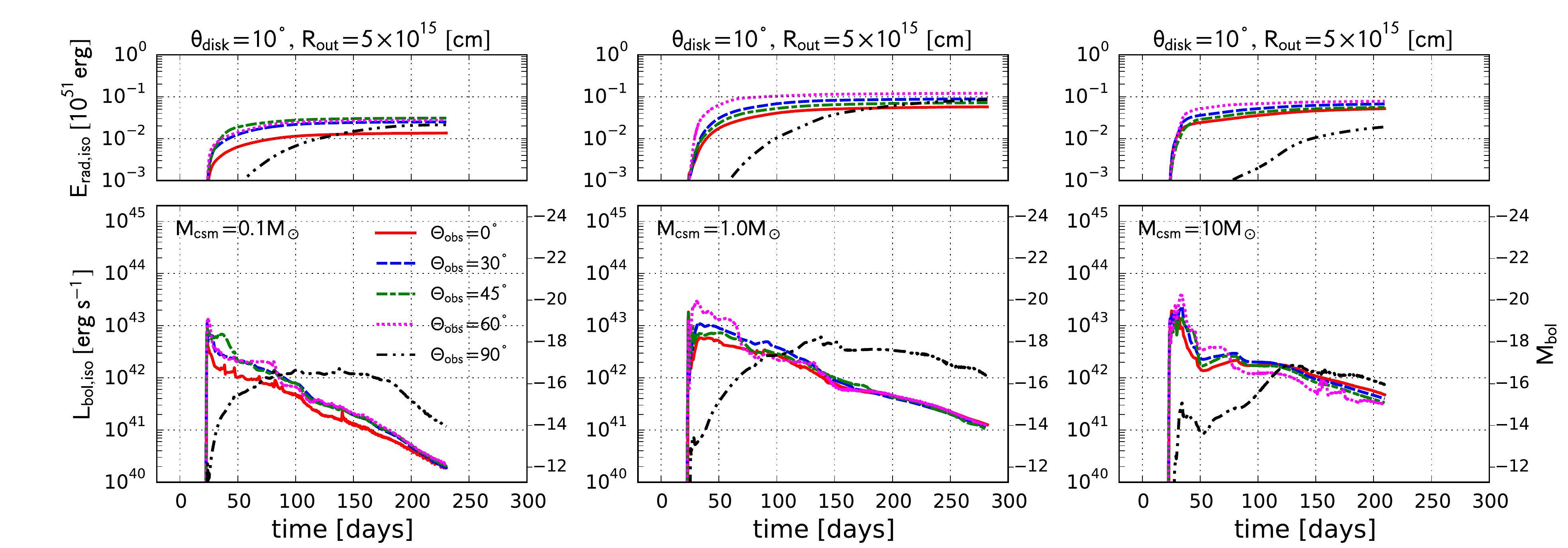}
\cprotect\caption{Isotropic equivalent bolometric light curves of the disk CSM models with $\theta_\mathrm{disk}=10^\circ$ and $R_\mathrm{out}=5\times 10^{15}$ cm. 
From left to right, we plot the models with the CSM mass of $0.1$, $1.0$, and $10M_\odot$ (models \verb|D10_M01|, \verb|D10_M1|, and \verb|D10_M10|). 
The light curves with viewing angles of $0^\circ$, $30^\circ$, $45^\circ$, $60^\circ$, and $90^\circ$ are plotted. 
In the upper panel, the cumulative radiated energy is also plotted for each viewing angle. }
\label{fig:lc_d10}
\end{center}
\end{figure*}

\begin{figure*}
\begin{center}
\includegraphics[scale=0.46]{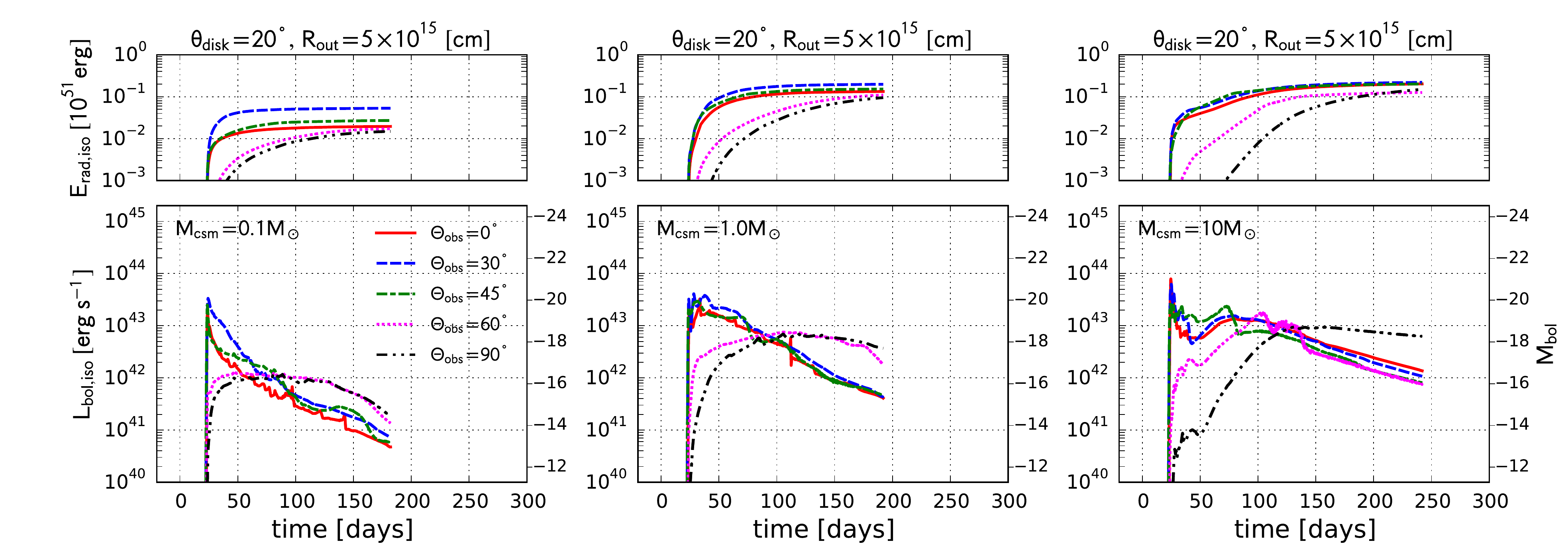}
\cprotect\caption{Same as Figure \ref{fig:lc_d10}, but for model \verb|D20_M1|. 
}
\label{fig:lc_d20}
\end{center}
\end{figure*}

In Figures \ref{fig:lc_d10} and \ref{fig:lc_d20}, we plot the isotropic-equivalent bolometric light curves for the disk CSM models. 
The light curves with different viewing angles show marked differences from those of the spherical CSM models with the same CSM mass (Figure \ref{fig:lc_sph}). 
In addition, some viewing angle effects are clearly recognized. 
We can divide the presented light curves into two classes, fast and slow risers. 
The light curves with smaller viewing angles are characterized by a fast rise. 
On the other hand, an observer around the equator ($\Theta_\mathrm{obs}=90^\circ$) would see less luminous and slowly evolving emission lasting for $\sim 1$ yr. 

The steep rise seen in light curves with smaller viewing angles can be explained by emission escaping into the regions that are not covered by the CSM disk. 
Without high density material preventing radiation from escaping, the observer can see the photosphere located in the SN ejecta directly. 
The radiation energy initially attached to the SN ejecta is simply released into interstellar space as the ejecta expand. 
However, the initial radiation energy in the SN ejecta alone cannot explain the luminosity at the initial peak. 
Therefore, the ejecta--CSM interaction happening around the equator also contributes to the emission along small viewing angles. 
In other word, the energy dissipated at the ejecta--CSM interface diffuse throughout the SN ejecta and then released into regions that are not covered by the CSM disk.

Among light curves in each panel of Figures \ref{fig:lc_d10} and \ref{fig:lc_d20}, those with larger viewing angles ($\Theta_\mathrm{obs}=90^\circ$ for Figure \ref{fig:lc_d10}  and $\Theta_\mathrm{obs}=60^\circ$ and $90^\circ$ for Figure \ref{fig:lc_d20}) are classified into the slowly evolving case. 
These light curves exhibit a slow rise followed by a slow decay. 
As seen in the density distributions in Figures \ref{fig:snap_disk1} and \ref{fig:snap_disk2}, the lines of sight corresponding to these viewing angles are intercepted by the CSM. 
Therefore, the observer sees the emission diffusing through the CSM disk.

\begin{figure*}
\begin{center}
\includegraphics[scale=0.46]{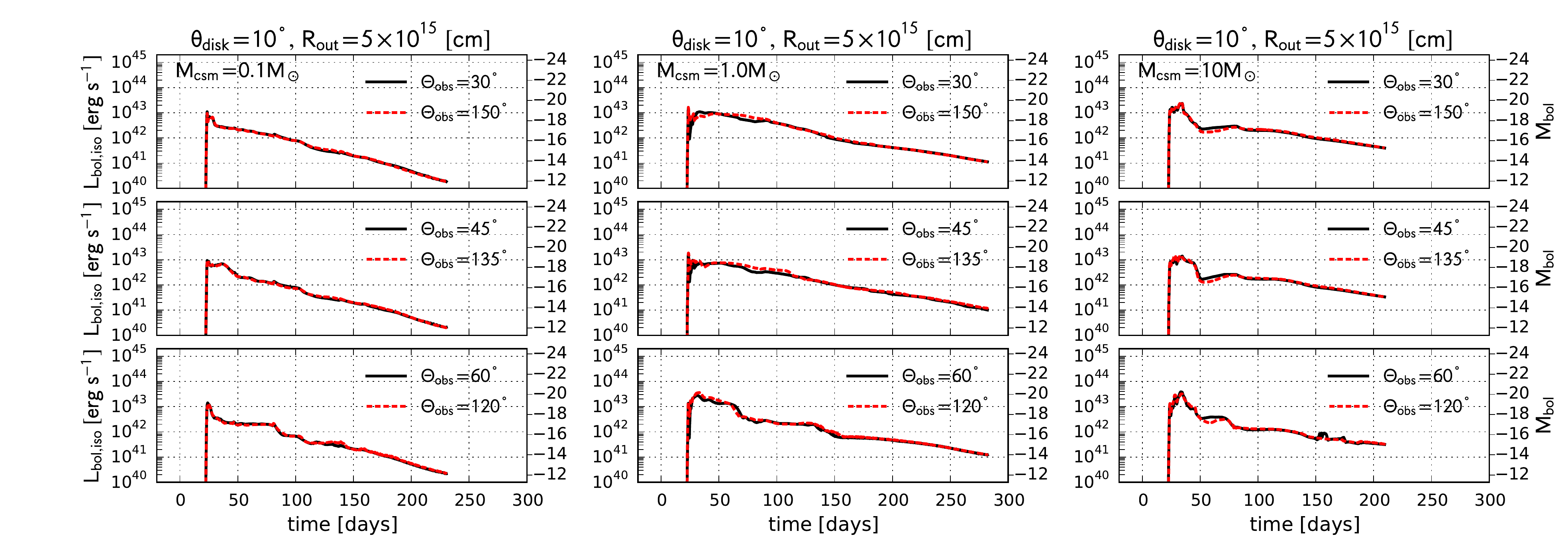}
\cprotect\caption{Comparison of light curves with $\Theta_\mathrm{obs}=30^\circ$ and $150^\circ$, $45^\circ$ and $135^\circ$, and $60^\circ$ and $120^\circ$ for models \verb|D10_M01| (left), \verb|D10_M1| (center), and \verb|D10_M1| (right).}
\label{fig:lc_ul}
\end{center}
\end{figure*}

As seen in Figures \ref{fig:snap_disk1} and \ref{fig:snap_disk2}, the shocked ejecta and CSM do not show strict equatorial symmetry. 
As a result, light curves with a viewing angle $\Theta_\mathrm{obs}=\Theta$ and its symmetric counterpart $\Theta_\mathrm{obs}=180^\circ-\Theta$ are not identical with each other. 
In Figure \ref{fig:lc_ul}, we compare light curves with $\Theta_\mathrm{obs}=30^\circ$ and $150^\circ$, $45^\circ$ and $135^\circ$, and $60^\circ$ and $120^\circ$ for models with $\vartheta_\mathrm{disk}=10^\circ$. 
Although each pair of light curves show slight differences, they mostly behave in a similar way. 
We have also checked the light curves of the models with $\vartheta_\mathrm{disk}=20^\circ$ and confirmed that the effect of the equatorial asymmetry on light curves is less than a factor of 2.

\subsubsection{Non-monotonic luminosity evolution\label{sec:non_monotonic}}
\begin{figure*}
\begin{center}
\includegraphics[scale=0.8]{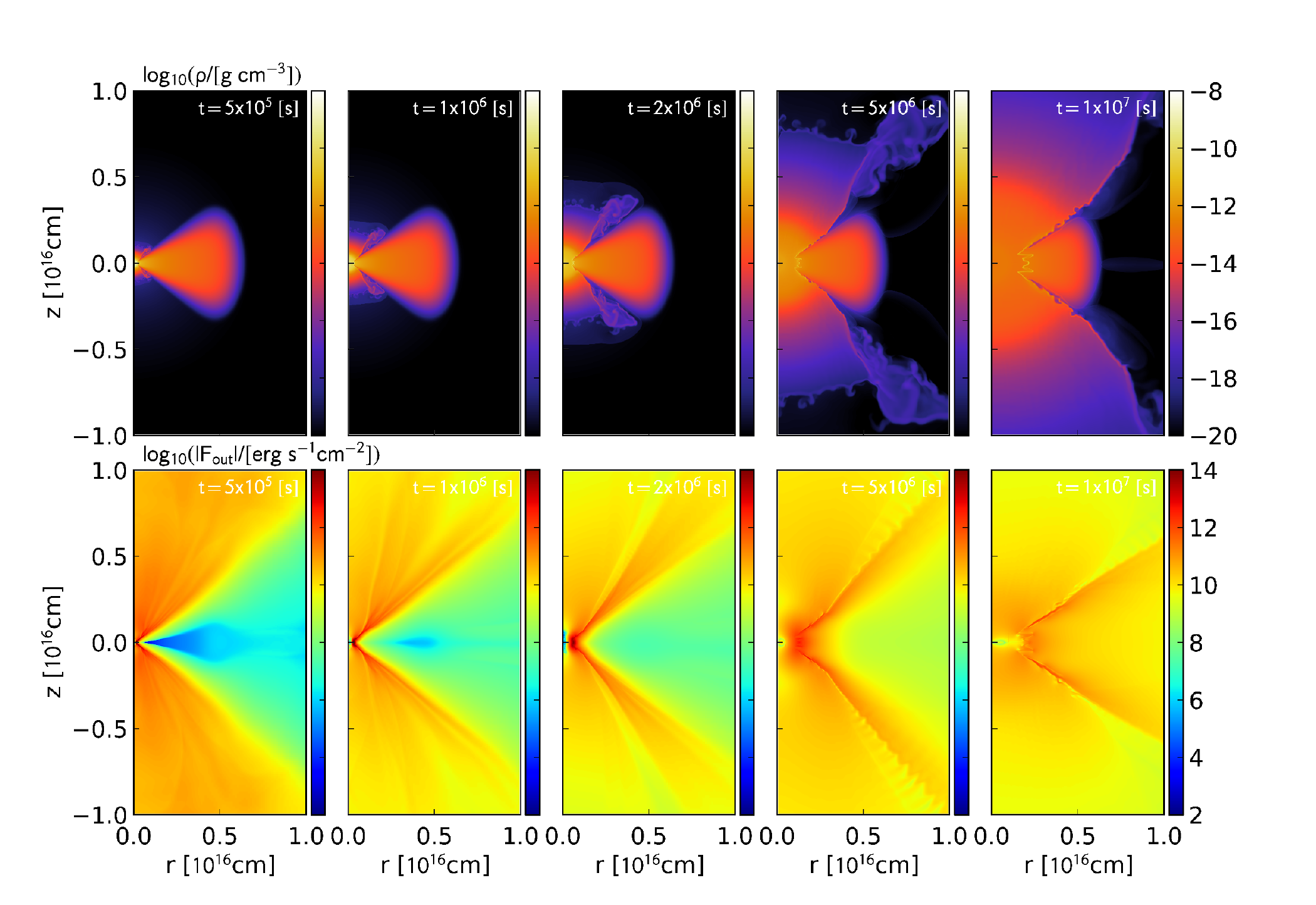}
\cprotect\caption{Spatial distributions of the density (upper) and the outgoing flux (lower) for model \verb|D20_M10|. 
Snapshots at $t=5\times 10^5$, $10^6$, $2\times10^6$, $5\times 10^6$, $10^7$ s are presented from left to right.}
\label{fig:flux}
\end{center}
\end{figure*}

Some light curves in Figures \ref{fig:lc_d10} and \ref{fig:lc_d20} show non-monotonic temporal evolution. 
The light curves appear to be more bumpy for a larger CSM mass. 
This is a result of the complex hydrodynamic interaction between the SN ejecta and the inner disk. 
We consider model \verb|D20_M10|, which shows the widest variety of light curves among the disk models.  
Figure \ref{fig:flux} shows the spatial distributions of the density and the outgoing flux at several epochs for model \verb|D20_M10|. 
The outgoing flux is calculated by Equation (\ref{eq:f_out}) with $l_\mathrm{v}^i=(r/R,z/R)$ at each numerical cell. 
These outgoing flux distributions reflect the distributions of the density and the radiation energy source at the corresponding epoch. At early epochs, the ejecta are interacting with an inner and denser part of the disk and thus the disk region is not penetrated by radiation, leading to small outgoing fluxes. 
The outgoing flux is larger around the region with lower inclination angles, which is not covered by the disk. 
Furthermore, at the earliest epoch ($t=5\times 10^5$ s), even regions close to the symmetry axis, $z=0$, show high outgoing fluxes, which indicates that the radiation produced by the interaction region can easily escape into directions with smaller viewing angles. 
At the early stages of the ejecta-disk interaction, the energy dissipation happens at outer layers of the ejecta. 
Therefore, the relatively small optical depth of the outer layers makes it easy for the dissipated energy to radiate away into wider solid angles including inclination angles close to $0^\circ$. 
Then, the interaction region digs through inner regions of the ejecta (in mass coordinate) and serves as an energy source nearly at the center. 
Therefore, for ejecta with smaller inclination angles, the energy dissipated at the deeply embedded interaction region should diffuse through the stratified layers.  
These two effects explain the light curves of \verb|D20_M10| with $\Theta_\mathrm{obs}=0^\circ$ and $30^\circ$ (Figure \ref{fig:lc_d20}). 
They exhibit an initial spike followed by a slowly evolving second peak. 
The initial spike is created by the impact of the outer ejecta colliding with the inner disk, which can easily escape into smaller viewing angles. 
The second peak is powered by photons diffusing from the deeply embedded energy dissipation region all the way toward the outermost layer. 

The light curve with $\Theta_\mathrm{obs}=45^\circ$ behaves in a more complicated way because its viewing angle is close to the orientation of the ejecta-disk interface. 
As seen in Figure \ref{fig:flux}, the region with the highest outgoing flux is aligned with the upper and lower disk surfaces. 
This is naturally expected for the ejecta-disk interaction where the energy dissipation happens most efficiently in the inner disk edge. 
Above (below) the upper (lower) disk surface with a large outgoing flux, however, regions with lower outgoing fluxes can be found. 
These regions are associated with the ejecta dynamically influenced by the collision with the inner disk. 
In the ejecta-disk collision, the ejecta initially expanding into equatorial direction are pushed away by the inner disk digging through the inner part of the ejecta, and accumulate above (below) the upper (lower) disk surface. 
The resultant region with an enhanced density around the disk surface can be seen in the density distributions in the upper panels of Figure \ref{fig:flux}. 
In these regions, the boosted optical depth along the radial direction creates ``shadowed regions'' (the regions with relatively lower $F_\mathrm{out}$ extending along $\theta=30^\circ$--$45^\circ$ and $135^\circ$--$150^\circ$) in the outgoing flux distribution, leading to striped outgoing flux distributions as seen in the lower panels of Figure \ref{fig:flux}. 
In addition, the orientation of the region with enhanced density can change with time due to the compression of the disk by the ejecta. 
Accordingly, the shadowed region is more deeply lied down on the disk surface at late epochs. 
It is how a particular line of sight intersects with striped flux distributions that determines the temporal behavior of the corresponding light curve. 
As a result of these effects, the light curves exhibit non-monotonic evolution. 

We also note that the outgoing flux distribution at $t=5\times 10^6$ s show some small-scale structure in the shadowed region. 
This region corresponds to clumpy ejecta along the ejecta-disk interface. 
Therefore, this structure may be related to the shear layer affected by the possible growth of the Kelvin-Helmholz instability, although we can not exclude the possibility of numerical oscillations. 
Since the typical length of this small-scale perturbation is less than $10^{15}$ cm, which yields a light crossing time of less than 1 day, this behavior does not produce 10 days-long bumps observed in light curves even when it is really physical phenomena.

\begin{figure*}
\begin{center}
\includegraphics[scale=0.55]{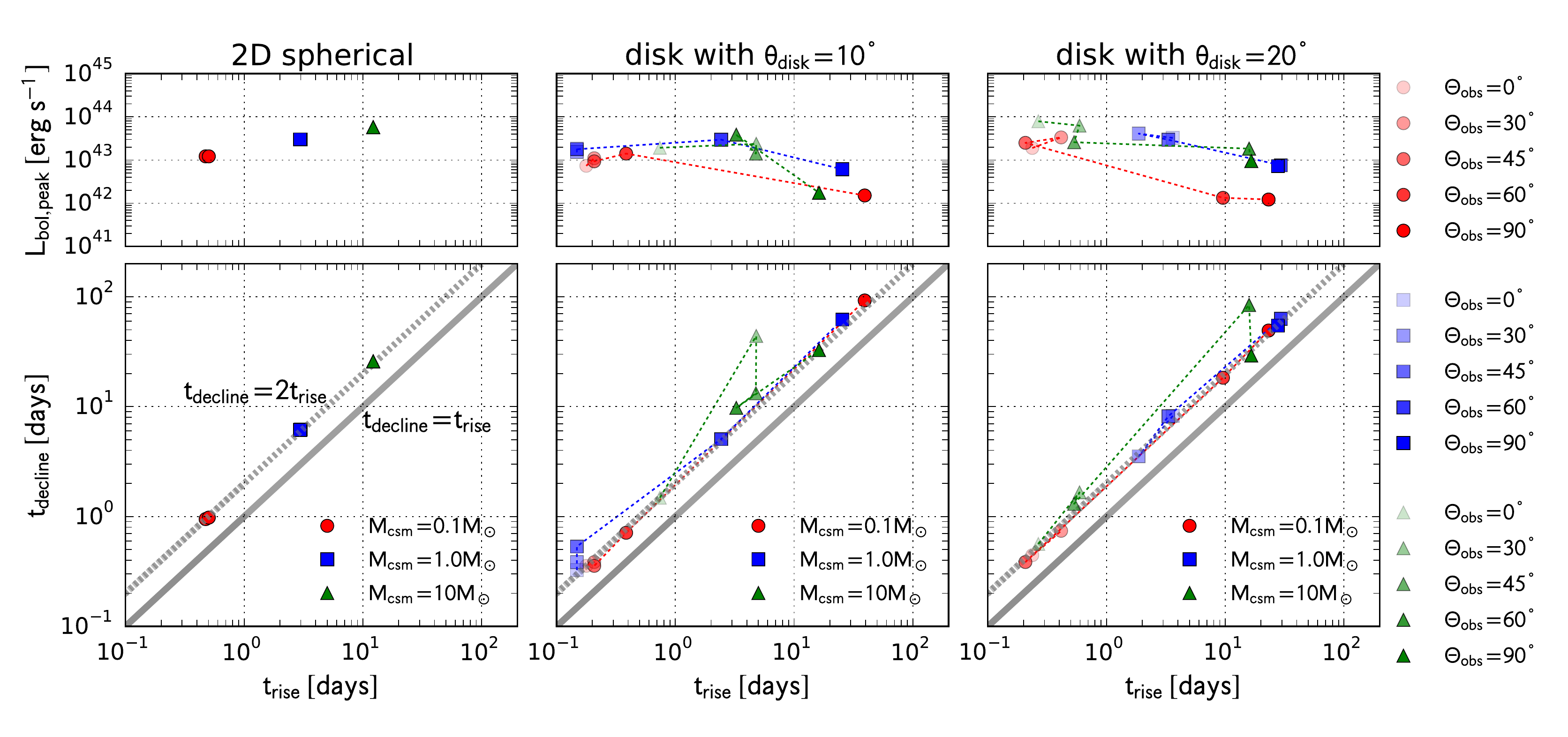}
\cprotect\caption{Peak bolometric luminosity and decline timescale as a function of rise timescale. 
}
\label{fig:rise_time}
\end{center}
\end{figure*}

\subsection{Rise and decline times\label{sec:rise_time}}
For a more quantitatively look at the light curves, we introduce the rise and decline timescales. 
How a transient population behaves in the duration--luminosity phase space can give us information on the emission mechanism \citep[e.g.,][]{2014ApJ...788..154O,2014ApJ...790L..16M} and also potentially discriminate among different transient populations \citep{2017ApJ...849...70V}. 
Several observational studies on rapid transients introduce the time above the half-maximum \citep[e.g.,][]{2014ApJ...794...23D}, the inverse of a magnitude change per unit time \citep[e.g.,][]{2016ApJ...819....5T}, or rising timescale based on a polynomial fit \citep[e.g.,][]{2016ApJ...819...35A}. 
These methods are based on the measurement of the slope of a light curve around the maximum or the fitting of a simple function to observed light curves.
Since some of our light curves show several humps, these methods may only provide the characteristic timescale of a single hump at the maximum light and may not capture the overall evolutionary trend of a light curve. 
Therefore, we use the cumulative radiated energy, (Equation \ref{eq:E_rad}), instead of light curve shapes. 
First, for a given light curve, the peak time $t_\mathrm{peak}$ is simply defined as the time at which the luminosity reaches the maximum. 
The radiated energy at $t=t_\mathrm{peak}$ is denoted by $E_\mathrm{rad,peak}(\Theta_\mathrm{obs})=E_\mathrm{rad,iso}(\Theta_\mathrm{obs},t_\mathrm{peak})$. 
Then, the rise time $t_\mathrm{rise}$ is defined as the time during which the half of the peak radiated energy is emitted before $t=t_\mathrm{peak}$;
\begin{equation}
E_\mathrm{rad,iso}(\Theta_\mathrm{obs},t_\mathrm{peak}-t_\mathrm{rise})=
0.5E_\mathrm{rad,peak}(\Theta_\mathrm{obs}).
\end{equation}
In a similar way, the decline time $t_\mathrm{decline}$ is defined as the time during which the same amount of the peak radiated energy is emitted after $t=t_\mathrm{peak}$;
\begin{equation}
E_\mathrm{rad,iso}(\Theta_\mathrm{obs},t_\mathrm{peak}+t_\mathrm{decline})=
2E_\mathrm{rad,peak}(\Theta_\mathrm{obs}).
\end{equation}
We should note that our timescale calculation is based on bolometric light curves. Therefore, the timescales shown below would probably be different from those based on single-band light curves. 
In particular, given that the early emission is likely dominated by UV photons, light curves in an optical or infrared band would rise more slowly than the bolometric counterparts. 
With this potential systematic difference in mind, we calculated the rise and decline times for the light curves shown in Figures \ref{fig:lc_sph}, \ref{fig:lc_d10}, and \ref{fig:lc_d20}. 

In Figure \ref{fig:rise_time}, we plot the results for the 2D spherical and two disk models in the  $t_\mathrm{rise}$--$L_\mathrm{peak}$ and $t_\mathrm{rise}$--$t_\mathrm{decline}$ diagrams. 
A general trend in the rise time vs decline time plots (lower panels) is that rise and decline times are distributed around the dotted line showing $t_\mathrm{decline}=2t_\mathrm{rise}$, i.e., decline times are about two times longer than the corresponding rise times. 
For disk CSM models, some models show decline timescales a bit longer than $2t_\mathrm{rise}$, which qualitatively confirms the rapid rise followed by slowly decaying light curves.  
As we have discussed in Section \ref{sec:non_monotonic}, disk models exhibit complicated temporal evolution of the bolometric luminosity. 
For smaller viewing angles, the rising part of the light curves should be powered by the impact of the outer part of the ejecta on the inner edge of the disk. 
At this phase, the timescale of photon diffusion is determined by the outer and dilute part of the ejecta. 
At late epochs, the interaction region is located in the inner part of the ejecta and serves as an energy source around the central region, where the diffusion timescale could be longer than those at outer layers. 
This effect possibly produces some outliers in the $t_\mathrm{decline}$--$t_\mathrm{rise}$ plot, especially for larger CSM masses, which are more capable of penetrating the expanding ejecta. 
Again, light curves of disk models are strongly dependent on the viewing angle, while those of 2D spherical models are not. 
When the ejecta-disk interaction is seen through the disk CSM around the equator, the photon diffusion effect smears out the light curve, thereby making the rise and decline times longer. 
The prolonged rise and decline times still roughly follow the $t_\mathrm{decline}=2t_\mathrm{rise}$ relation, which makes it difficult to distinguish whether these timescales are prolonged due to the increase in the total CSM mass or the concentration of the CSM around the equator. 
On the other hand, the peak luminosity vs rise time plots (upper panels) behave differently. 
For 2D spherical models, the peak luminosity monotonically increases for increasing CSM mass. 
This is due to the increased efficiency for the dissipation of the ejecta kinetic energy. 
For disk models, larger viewing angles lead to lower peak luminosity and longer rise times. 
A CSM concentrated around the equator makes light curves significantly stretched. 
Even though the timescales are prolonged, the total dissipated energy is similar for given ejecta and CSM properties. 
The ejecta-CSM interaction happens in a deeply embedded region in the ejecta and the subsequent photon diffusion through the ejecta and CSM makes the radiation field nearly isotropic. 
As a result, a similar amount of radiation energy is distributed into different solid angles. 
Thus, the prolonged rise and decline times lead to lower peak luminosities as seen in the luminosity--rise time plot. 

These diagrams could be used to statistically infer the cause of the variety of light curves of interacting SNe by comparisons with observed samples. 
In the $t_\mathrm{decline}$--$t_\mathrm{rise}$ plot, both increasing the CSM mass and observing at larger viewing angles make the timescales longer. 
In the $L_\mathrm{peak}$--$t_\mathrm{rise}$ plot, on the other hand, large CSM masses make peak luminosities higher, while events with larger viewing angles are less luminous. 
For example, if the variety of the evolutionary timescales of SN IIn light curves are predominantly caused by the viewing angle effect with similar CSM masses, a declining trend could be found in the $L_\mathrm{peak}$--$t_\mathrm{rise}$ plot for type IIn SN samples. 
On the other hand, if the viewing angle dependence is only a minor effect and it is the CSM mass that predominantly determines evolution timescales, an increasing trend is found in the $L_\mathrm{peak}$--$t_\mathrm{rise}$ plot. 
Such hypotheses could be examined by using unbiased type IIn SN samples with well-measured $L_\mathrm{peak}$, $t_\mathrm{rise}$, and $t_\mathrm{decline}$. 
We briefly discuss this point with the type IIn SN samples compiled by \cite{2019arXiv190605812N} in Section \ref{sec:unveiling_csm_disks}.

Finally, we make a remark on the $t_\mathrm{rise}$--$t_\mathrm{decline}$ relation. 
As we have already noted, $t_\mathrm{decline}=2t_\mathrm{rise}$ relation appears to hold very well. 
The reason why this scaling relation holds well is unclear. 
It may reflect the self-similarity of the system \citep{1982ApJ...258..790C}. 
However, as opposed to purely hydrodynamic cases, the radiative diffusion effect introduces another characteristic length scale, making the situation more complicated. 
Although this relation is intriguing, it warrants further investigation and we regard the systematic investigation as one of the future works.

\section{Post-process calculations\label{sec:post_process}}
The properties of the thermal emission can be inferred from physical variables at the photosphere. 
In order to locate the photosphere and estimate the color temperature from the snapshots of the simulations, we carry out post-process calculations. 
The detailed numerical procedures are described in Appendix \ref{sec:locating_photosphere}.

\begin{figure*}
\begin{center}
\includegraphics[scale=0.2]{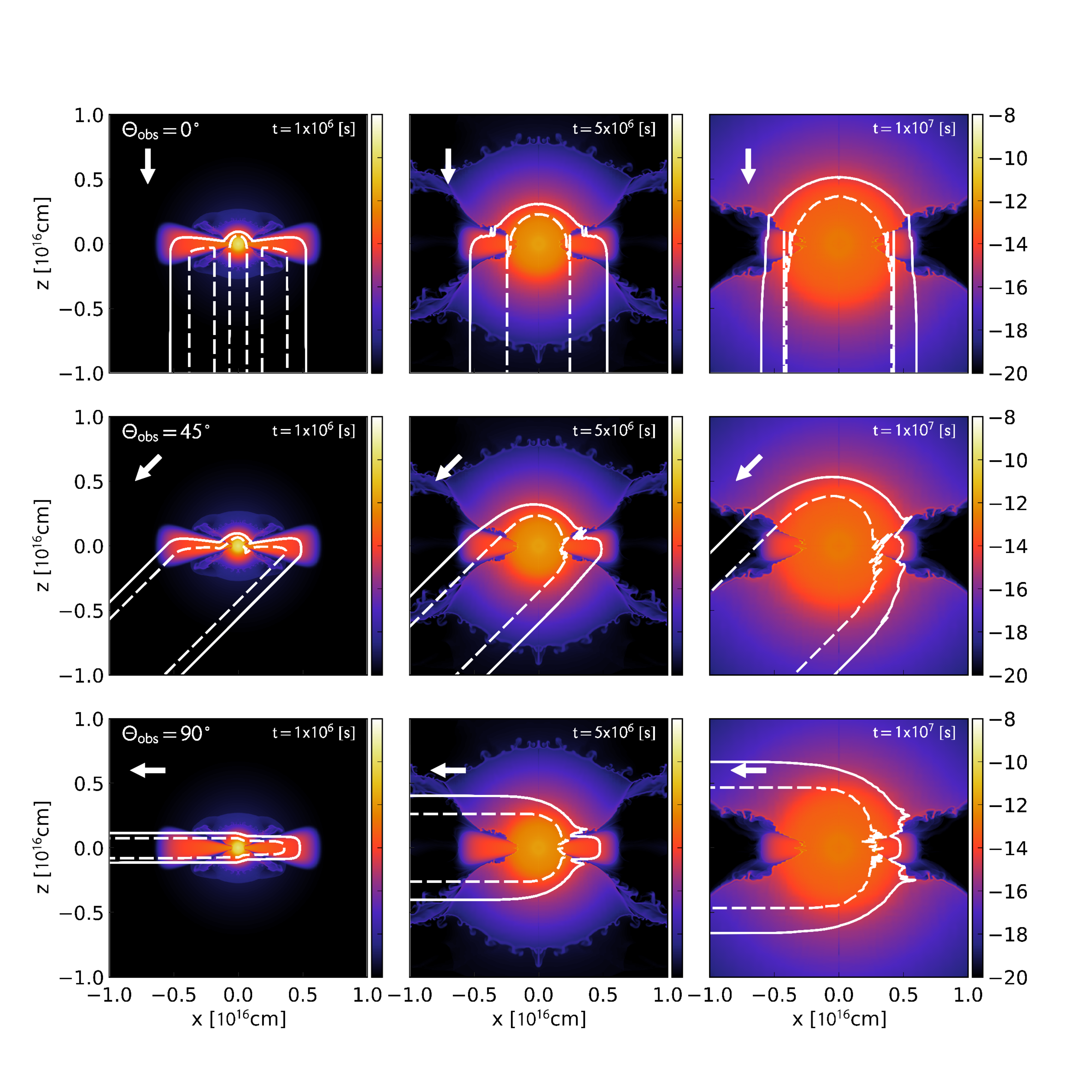}
\cprotect\caption{Locations of the photosphere from different viewing angles for model \verb|D10_M1|. 
In each row, the color-coded density distributions at $t=10^6$, $5\times10^6$, and $t=10^7$ s are presented from left to right. 
The solid and dashed curves (white) plotted over the density distribution show the locations where the optical depths defined by Equations (\ref{eq:tau}) and (\ref{eq:tau_eff}) give the threshold value of $2/3$.  
The viewing angle is set to $\Theta_\mathrm{obs}=0^\circ$ (top), $45^\circ$ (middle), and $90^\circ$ (bottom).
The line of sight is shown as the white arrow in each panel. 
}
\label{fig:photosphere}
\end{center}
\end{figure*}

We have estimated the color temperature and the blackbody radius for all the simulations and the viewing angles of $0^\circ$, $30^\circ$, $45^\circ$, $60^\circ$, and $90^\circ$. 
We have used up to $19$ snapshots of the simulations at elapsed times from $t=10^6$ to $1.9\times 10^7$ s (some models are terminated earlier). 
Figure \ref{fig:photosphere} shows some examples of the photosphere plotted over the density distribution. 
The results for model \verb|D10_M1| are presented. 
The scattering photosphere defined by $\tau=2/3$ (Equation \ref{eq:tau}) is located outside the effective photosphere $\tau_\mathrm{eff}=2/3$ (Equation \ref{eq:tau_eff}), suggesting that the photon production region is more deeply embedded in the SN ejecta. 
The geometry of the SN ejecta and the CSM disk apparently plays an important role in determining the photosphere. 
When the radius of the ejecta is smaller than the outer radius of the CSM at early epochs (e.g., the left panels of Figure \ref{fig:photosphere}), the CSM disk can be seen separated from the SN ejecta. 
At later epochs, however, a considerable fraction of the CSM disk is covered by the ejecta and thus the emission through the CSM can be seen only for observers around the equatorial plane. 
For observers with small viewing angles, the CSM disk would be hidden by the SN ejecta, while the ejecta-CSM interaction still contributes to the emission from the photosphere in the expanding ejecta. 
This geometrical effect on spectral evolution is further discussed below.

\begin{figure*}
\begin{center}
\includegraphics[scale=0.55]{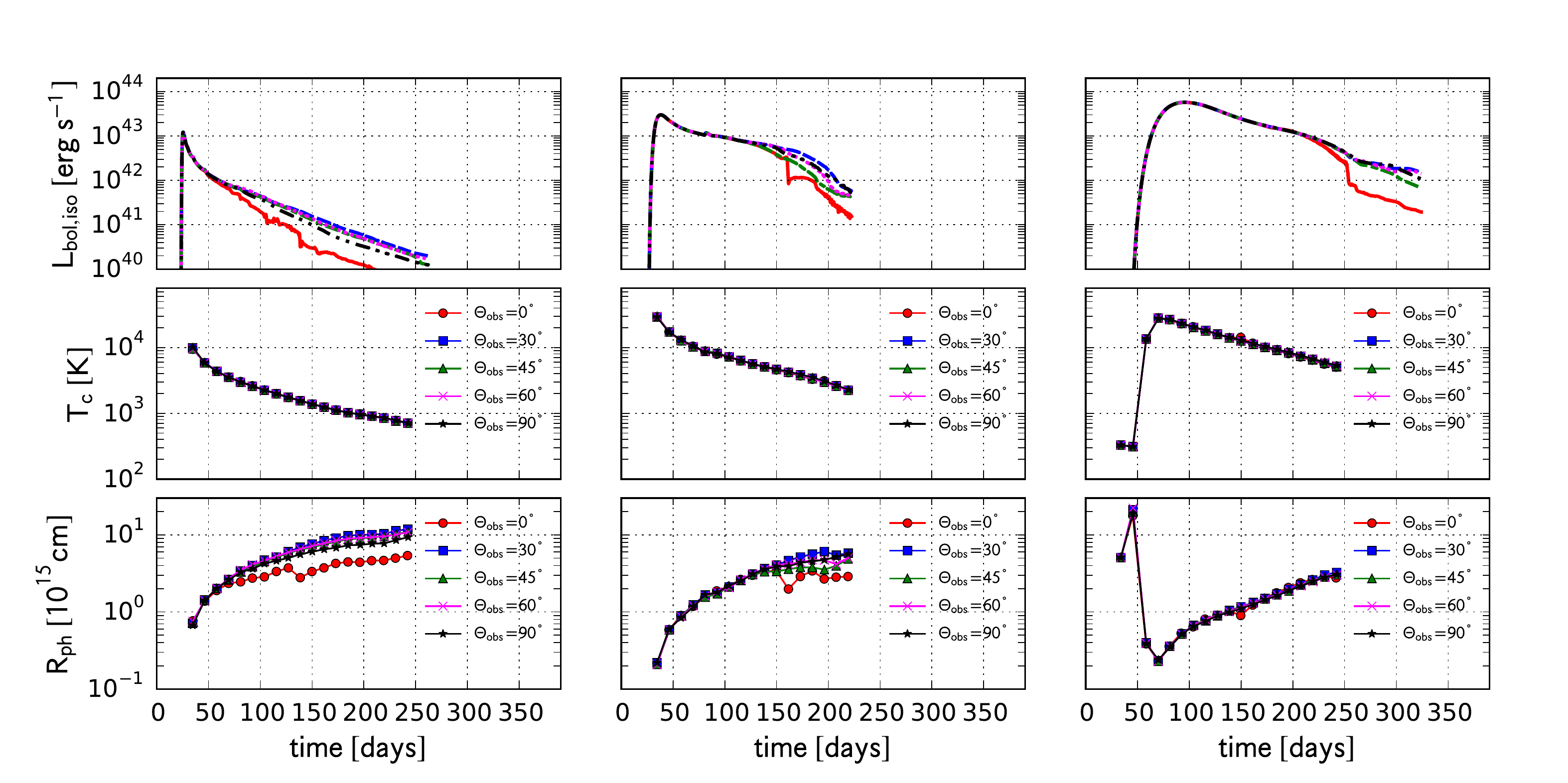}
\cprotect\caption{Color temperature and effective blackbody radius evolution for models \verb|S_M01| (left), \verb|S_M1| (center), and \verb|S_M10| (right). 
The top panels show the same bolometric light curves as in Figure \ref{fig:lc_sph}. 
In the middle and bottom panels, the color temperature and the blackbody radius are plotted as a function of time. 
The viewing angles are $\Theta_\mathrm{obs}=0^\circ$ (red circle), $30^\circ$ (blue square), $45^\circ$ (green triangle), $60^\circ$ (magenta cross), and $90^\circ$ (black star). 
}
\label{fig:color_sph}
\end{center}
\end{figure*}

\begin{figure*}
\begin{center}
\includegraphics[scale=0.55]{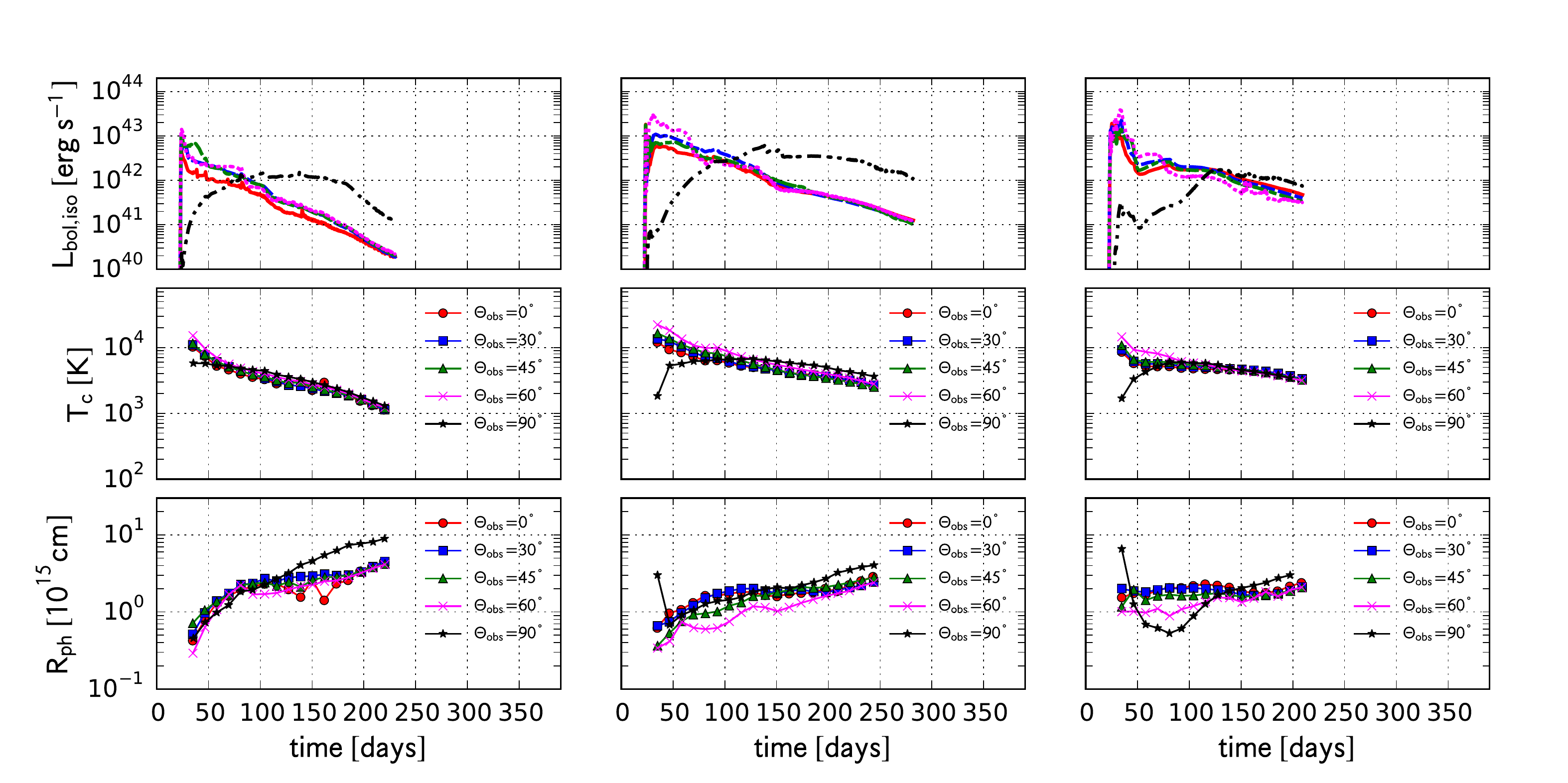}
\cprotect\caption{Same as Figure \ref{fig:color_sph}, but for models \verb|D10_M01|, \verb|D10_M1|, and \verb|D10_M10|. 
}
\label{fig:color_d10}
\end{center}
\end{figure*}

\begin{figure*}
\begin{center}
\includegraphics[scale=0.55]{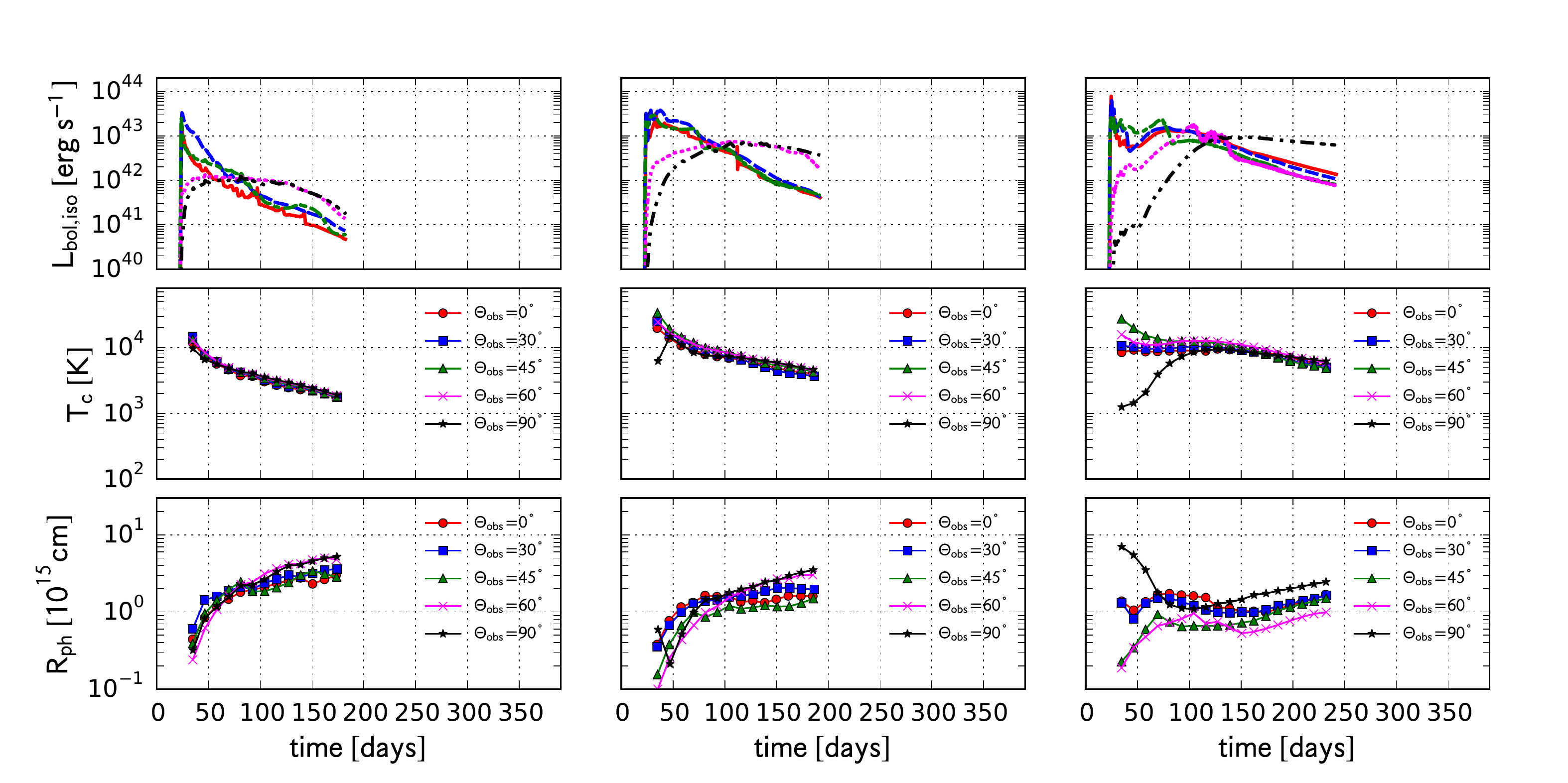}
\cprotect\caption{Same as Figure \ref{fig:color_sph}, but for models \verb|D20_M01|, \verb|D20_M1|, and \verb|D20_M10|. 
}
\label{fig:color_d20}
\end{center}
\end{figure*}

Figures \ref{fig:color_sph}, \ref{fig:color_d10}, and \ref{fig:color_d20} show the temporal evolutions of the color temperature and the effective blackbody radius for the 9 models. 
In general, the color temperature is initially of the order of $10^4$ K and then steadily decreases down to several $10^3$ K or lower. 
The blackbody radius generally increases with time. 
The color temperatures as high as $T_\mathrm{c,av}\sim 10^4$ K are in agreement with the photospheric temperature inferred by early observations of type IIn SNe. 
The steadily increasing blackbody radius is inconsistent with some type IIn SNe with detailed follow-up observations, e.g., SN 2006gy \citep{2010ApJ...709..856S}, whose blackbody radius increases in earlier epochs and then declines after entering the nebular stage. 
This discrepancy is probably due to the treatment of the opacity. 
We assume a constant opacity for electron scattering (Equation \ref{eq:kappa_es}), which implicitly assumes fully ionized gas. 
At later stages, however, the local gas temperature around the photosphere drops down to $\sim 6\times 10^3$ K and then free electrons start recombining, leading to a reduced electron scattering opacity. 
This is how the ejecta become transparent in hydrogen-rich SNe \citep[see, e.g.,][]{1996snih.book.....A}. 
Several models indeed show color temperatures lower than $\sim 6\times10^3$ K at later epochs, suggesting that the photosphere should be more compact. 

The color temperature decreases more slowly in models with larger CSM masses. 
On the other hand, the blackbody radii are generally smaller for models with more massive CSMs. 
A massive CSM makes the ejecta-CSM interaction long-lasting and thus the ejecta are kept hot. 
This can also make the SN ejecta more deeply embedded in the CSM as shown in Figure \ref{fig:shock}, keeping the blackbody radius small. 
For models with spherical CSMs, the color temperature evolutions do not strongly depend on the viewing angle. 
Although the blackbody radii exhibit slight variations, it is due to the late-time variety in the bolometric light curve. 

We also note that the effective blackbody radii $R_\mathrm{eff}$ for disk CSM models change with time modestly compared with spherical CSM models, ending up smaller values.  
Since $R_\mathrm{eff}$ is connected to the bolometric luminosity and the color temperature by the Stefan-Boltzmann law (Equation \ref{eq:stefan_boltzmann}), this behavior indicates that the color temperature is kept high even at late phases with low bolometric luminosity in the case of ejecta-disk interaction. 
This is because the inner disk edge penetrates the ejecta deeply and serves as a heating source for the expanding outer ejecta. 
However, the photospheric properties at late epochs also suffer from our simple treatment of opacity. 
Especially, neglecting hydrogen recombination may significantly affect the late photospheric properties, making the estimate of the color temperature and the blackbody radius uncertain. 


\section{Discussion}\label{sec:discussion}
In the following, we discuss observational signatures of SNe interacting with CSM disks based on the simulation results. 

\subsection{Photometric properties}
First of all, the 2D simulations with spherical CSMs presented in Section \ref{sec:spherical_csm} suggest that hydrodynamic instabilities developing in the ejecta-CSM interface have a very limited impact on bolometric light curves.  
This is because the energy dissipation region is deeply embedded in the dense CSM. 
On the other hand, the light curve of an SN interacting with a CSM disk strongly depends on the viewing angle. 
As we have seen in Section \ref{sec:disk_csm}, an observer with a smaller viewing angle sees the emission directly from the photohsphere in the expanding ejecta. 
This leads to steeply rising light curves. 
On the other hand, for observers with the line of sight intercepted by the CSM disk, photons created by the ejecta-disk interaction should diffuse through the CSM disk, making the rise and decline times longer. 
As our simulations suggest, whether a former or latter type of light curve is observed is roughly determined by the opening angle of the CSM disk. 
A CSM disk with a larger opening angle prevents a larger fraction of the ejecta from expanding freely, leading to an efficient dissipation of the original kinetic energy of the ejecta. 
Therefore, the opening angle of the CSM disk is one of the most fundamental parameters governing the dynamical evolution of the SN ejecta interacting with massive CSM disks. 

Aspherical CSM structure can also make the bolometric light curve much more complex than the spherical counterpart. 
For example, the bolometric light curves of model \verb|D20_M10| (Figure \ref{fig:lc_d20}) clearly exhibit several bumps. 
This feature is most prominent for viewing angles close to the boundary distinguishing the void region from the almost freely expanding ejecta and thus is likely produced as a consequence of the hydrodynamic interaction between the ejecta and the CSM disk. 
Interestingly, these bumpy light curves are reminiscent of the type IIn SN iPTF13z \citep{2017A&A...605A...6N} and the unusual transient iPTF14hls \citep{2017Natur.551..210A,2019A&A...621A..30S}. 
Remarkably, the latter has been bright for more than $1000$ days. 
The CSM interaction is one of the promising scenarios for explaining the extremely long and bright emission of this peculiar object \citep[e.g.,][]{2018MNRAS.477...74A,2018AstL...44..370C,2018ApJ...863..105W}. 
Although the timescale of the emission from iPTF14hls is much longer than that covered by our simulations, our simulation results may imply that the bumpy light curves with an exceptionally long duration could be explained by an SN interacting with a more extended CSM disk as hinted by the late-time H$\alpha$ line profile \citep{2018MNRAS.477...74A}. 
Although the ejecta-disk interaction is one possible way to produce light curves with several bumps as we have demonstrated by our simulations, it would not be the only way. 
Mass-loss processes prior to the gravitational collapse of massive stars might be sporadic. 
Therefore, multiple mass ejection episodes may produce quasi-spherical, but radially inhomogeneous CSM, which also likely leads to bumpy light curves. 

Very recently, \cite{2019arXiv190605812N} presented type IIn SN samples from the PTF survey. 
They report that the fraction of type IIn SNe similar to iPTF13z would be only $1.4^{+14.6}_{-1.0}\%$ of the whole type IIn SN population. 
This indicates that interacting SNe with bumpy light curves are relatively rare and some special condition may be required to produce a highly aspherical CSM. 
Even if CSM disks are common, only observers at the direction of the disk opening angle see the emission with bumpy light curves caused by the ejecta-disk interaction. 

\subsection{Emission lines}
We briefly comment on the emission line profile expected from SN ejecta interacting with a CSM disk. 

The asymmetric ejecta structure realized in our numerical simulations would have several impacts on the emission line profiles, for both narrow and broad lines seen in type IIn SNe. 
As seen in the top panels of Figure \ref{fig:snap_sph_m01}, the presence of a  sufficiently massive CSM disk around the SN ejecta decelerates the equatorial part of the ejecta. 
For example, from an observer seeing this event along the symmetry axis, the ejecta component traveling along the transverse direction to the line of sight is selectively decelerated. 
As a result, this component would contribute to broad emission lines less effectively, which may lead to a top hat line profile. 
On the other hand, for an observer standing around the equatorial plane, the ejecta components going toward and away from the observer are decelerated, which  possibly makes the blue- and red-shifted components in a line profile distorted significantly. 

For narrow emission and absorption line components originating from the CSM, the configuration of the SN ejecta and the CSM disk has even more drastic impacts on line shapes. 
As shown in Figure \ref{fig:photosphere}, the CSM disk is seen in early epochs when the radius of the SN ejecta is smaller than the outer radius of the CSM disk. 
However, the expanding SN ejecta gradually cover the CSM disk. 
The CSM disk would be hidden until the most part of the SN ejecta becomes transparent in the nebular stage. 
This geometrical effect suggests that the relative contributions of the narrow and broad lines, which are originated from the slowly-moving CSM and the fast SN ejecta, can exhibit complicated time variation. 
\cite{2015MNRAS.449.1876S} suggested this geometrical effect for the type IIn SN iPTF11iqb. 
Our results are qualitatively same as their scenario. 


A more quantitative discussion requires line transfer calculations by using snapshot of radiation-hydrodynamic simulations. 
Therefore, we leave such more detailed post-process computations one of our future work. 

\subsection{Unveiling CSM disks\label{sec:unveiling_csm_disks}}
The biggest question among SNe interacting with massive CSMs is how and when such mass ejection happened. 
On one hand, it may happen in unusual eruption events in a specific class of massive stars like LBVs. 
On the other hand, it may be a consequence of complicated interaction processes in close binary systems. 
Therefore, probing the CSM properties through SN observations is of great importance. 

Our radiation-hydrodynamic simulations have found that light curves of SNe colliding with CSM disks exhibit two distinct characteristics depending on viewing angles, a rapid rise followed by a steady decline and a slow rise followed by a slow decline. 
The former correspond to ejecta-disk interection with face-on geometry and the latter corresponds to edge-on geometry. 
Identifying and collecting these two types of SNe harbouring CSM disk interaction would reveal important properties of putative CSM disks associated with exploding massive stars. 
For example, the typical opening angle of the CSM disk could be probed by the relative fractions of interacting SNe with former and latter types of light curves. 

Statistical analyses of unbiased type IIn SN samples is one possible way to understand the variety of their light curve characteristics. 
In \cite{2019arXiv190605812N}, they provide the peak luminosity vs rise time plot for their PTF samples. 
The correlation between the rise time and the peak luminosity is not statistically significant. 
However, they found that luminous SNe IIn generally show longer evolution timescales. 
If this is the case, this observational trend is at odds with the trend produced by the viewing angle effect (see the upper panels of Figure \ref{fig:rise_time}). 
It would be the CSM mass that makes SNe IIn outshine more brightly and in longer timescales at the same time, although anisotropic CSM and the viewing angle effect should play a role in producing the dispersion in the trend.

\section{Conclusions}\label{sec:conclusions}
In this work, we have performed 1D spherical and 2D cylindrical radiation-hydrodynamic simulations of interacting SNe. 
For 2D simulations, we have considered SN ejecta interacting with spherical and disk-like CSMs. 

In the 2D spherical CSM models, the overall evolution of the ejecta interacting with the CSM is similar to the 1D spherical counterparts. 
Although hydrodynamic instabilities certainly develop at the ejecta-CSM interface, the filaments or clumps produced by the instability is confined in the narrow layer between the forward and reverse shock fronts and do not modify the global shape of the expanding ejecta. 
Consequently, the bolometric light curves with different viewing angles are similar to those of 1D spherical models. 

On the other hand, disk-like CSMs have much more impacts on the dynamics of the ejecta and the resultant light curves. 
A part of the ejecta traveling around the equator collide with the massive CSM, leading to the dissipation of the kinetic energy. 
The collision around the equator and the subsequent diffusion of the radiation throughout the CSM disk give rise to bright emission, which can be seen even for observers with small viewing angles. 

Finally, we note some remarks on this study. 
Our simulations treat frequency-integrated radiation energy density and radiative fluxes. 
Although the radiation pressure effect and the coupling between gas and radiation (in gray approximation) are included, information on the color temperature and the photospheric radius are not directly obtained from the simulations. 
Another potential caveat is the use of the two-temperature approximation. 
Although it is relatively easy to handle gas-radiation coupling under this approximation in multi-dimensional simulations, it is not always valid. 
Especially, for fast radiative shocks propagating in dilute media, the Compton scattering can be a dominant process realizing the gas-radiation equilibrium \citep{1976ApJS...32..233W}. 
The inefficient coupling between gas and radiation produces a wider relaxation layer with higher post-shock gas temperature, where high-energy photons in X-ray and even gamma-ray energy ranges could be created. 
This situation would be realized in a supernova shock breakout in relatively dilute wind \citep[e.g.,][]{2010ApJ...716..781K,2017hsn..book..967W}. 
In our approach, however, we simply treat electron scattering as elastic scattering in the comoving frame of the gas flow. 
In addition, we have assumed that the gas is fully ionized and the electron scattering and free-free absorption are major radiative processes. 
While these assumptions are valid in early epochs, when the interaction layer is kept hot by the energy dissipation, recombination effects and bound-free opacities become more and more important at later epochs, which would affect late-time light curves. 
These issues should be improved in future studies. 

Nevertheless, our simulations clearly demonstrate the interaction of SN ejecta with aspherical CSMs can lead to a wide variety of CCSNe interacting with their surrouding gas. 
Light curve and spectral modelings taking aspherical CSMs into account as well as scrutinizing well-observed type IIn SNe could be a key to elucidating the mysterious origin of massive CSMs in the immediate vicinity of massive stars.  

\acknowledgements
We appreciate the anonymous referee for his/her constructive comments on the manuscript. 
A.S. acknowledges support by Japan Society for the Promotion of Science (JSPS) KAKENHI Grand Number JP19K14770. 
This study was also supported in part by the Grants-in-Aid for the Scientific Research of Japan Society for the Promotion of Science (JSPS, Nos. 
JP17H02864, 
JP18K13585, 
JP17H01130, 
JP17K14306, 
JP18H01212  
),  
the Ministry of Education, Science and Culture of Japan (MEXT, Nos. 
JP17H06357, 
JP17H06364 
),
and by JICFuS as a priority issue to be tackled by using Post `K' Computer.
Numerical simulations were carried out by Cray XC50 system operated by Center for Computational Astrophysics, National Astronomical Observatory of Japan. 

\software{Matplotlib (v2.2.3; \citealt{Hunter:2007})
}

\appendix
\section{Numerical Integration of equations of radiation-hydrodynamics}\label{sec:numerical_integration_detail}
In this section, we describe the way to numerically integrate equations of radiation hydrodynamics, Equations (\ref{eq:eq_Erad})--(\ref{eq:eos}), (\ref{eq:G0}), and (\ref{eq:Gi}), which is updated from the previous code \citep{2016ApJ...825...92S}. 
As we have mentioned in Section (\ref{sec:equations_of_radiation_hydrodynamics}), the advection part of the equations are integrated in a standard way. 
Therefore, we focus on the source terms of the equations in the following.

\subsection{Implicit Integration of the source terms\label{sec:source_term_numerical}}
A simple discretization of the governing equations gives
\begin{equation}
E_\mathrm{r}-E_\mathrm{r}^{n}=G^{0}\Delta t,
\end{equation}
and
\begin{equation}
F_\mathrm{r}^{i}-F_\mathrm{r}^{i,n}=G^{i}\Delta t,
\end{equation}
where $E_\mathrm{r}^n$ and $F_\mathrm{r}^{i,n}$ are radiation energy density and flux at $t=t^n$, while $E_\mathrm{r}$ and $F_\mathrm{r}^{i}$ are those at the next time step, $t=t^n+\Delta t$. 
We evaluate the right-hand sides of the equations at $t=t^{n}+\Delta t$, i.e., we solve the equations implicitly. 

Some algebraic manipulations lead to the following convenient equations,
\begin{equation}
E_\mathrm{r}-E_\mathrm{r}^n-\beta_i\left(F_\mathrm{r}^{i}-F_\mathrm{r}^{i,n}\right)=
\frac{\bar{\rho}\bar{\kappa}_\mathrm{a}\Delta t}{\Gamma}
\left(a_\mathrm{r}\bar{T}_\mathrm{g}^4-\bar{E}_\mathrm{r}\right),
\end{equation}
and
\begin{equation}
-\beta^i\left(E_\mathrm{r}-E_\mathrm{r}^n\right)+F_\mathrm{r}^i-F_\mathrm{r}^{i,n}=
-\bar{\rho}(\bar{\kappa}_\mathrm{a}+\bar{\kappa}_\mathrm{s})\Delta t
\left(\delta^i_j-\frac{\Gamma}{\Gamma+1}\beta^i\beta_j\right)\bar{F}_\mathrm{r}^j,
\end{equation}
We note that the velocity $\beta^i$ and the gas temperature $\bar{T}_\mathrm{g}$ in these equations are values at $t=t^n+\Delta t$. 
The radiation energy density and the flux, $E_\mathrm{r}$ and $F_\mathrm{r}^i$, in the laboratory frame, appearing in the left-hand sides of these equations can be eliminated by using the Lorentz transformations, Equation (\ref{eq:transformation_Er}) and (\ref{eq:transformation_Fr}). 
Thus, one obtains
\begin{equation}
\bar{E}_\mathrm{r}+\beta_i\bar{F}_\mathrm{r}^i-E_\mathrm{r}^n+\beta_iF_\mathrm{r}^{i,n}=\Delta \tau_\mathrm{ab}\Gamma^{-2}
\left(a_\mathrm{r}\bar{T}_\mathrm{g}^4-\bar{E}_\mathrm{r}\right),
\end{equation}
and
\begin{equation}
\Gamma\left(
\delta^i_j-\frac{\Gamma}{\Gamma+1}\beta^i\beta_j\right)(\bar{F}_\mathrm{r}^j+\beta_k\bar{P}_\mathrm{r}^{jk})
-F_\mathrm{r}^{i,n}+\beta^iE_\mathrm{r}^n=
-\Delta\tau_\mathrm{as}\Gamma^{-1}\left(\delta^i_j-\frac{\Gamma}{\Gamma+1}\beta^i\beta_j\right)\bar{F}_\mathrm{r}^j,
\end{equation}
where
\begin{equation}
\Delta \tau_\mathrm{ab}=\bar{\rho}\Gamma\bar{\kappa}_\mathrm{a}\Delta t,
\end{equation}
and
\begin{equation}
\Delta \tau_\mathrm{as}=\bar{\rho}\Gamma(\bar{\kappa}_\mathrm{a}+\bar{\kappa}_\mathrm{s})\Delta t.
\end{equation}

Here we assume that the Eddington tensor $\bar{D}_\mathrm{r}^{ij}=\bar{P}_\mathrm{r}^{ij}/\bar{E}_\mathrm{r}$ can be approximately treated as a constant tensor during the time step from $t=t^n$ to $t=t^n+\Delta t$. 
Then, the above equations can be solved with respect to the comoving radiation energy density,
\begin{equation}
\bar{E}_\mathrm{r}=
\frac{\left(1+\Delta\tau_\mathrm{as}\Gamma^{-2}\right)
\left(E_\mathrm{r}^{n}-\beta_iF_\mathrm{r}^{i,n}+\Delta\tau_\mathrm{ab}\Gamma^{-2}a_\mathrm{r}\bar{T}_\mathrm{g}^4\right)
+\beta_i\left(E_\mathrm{r}^{n}\beta^{i}-F_\mathrm{r}^{i,n}\right)
}
{\left(1+\Delta\tau_\mathrm{ab}\Gamma^{-2}\right)\left(1+\Delta\tau_\mathrm{as}\Gamma^{-2}\right)-\beta_j\beta_k\bar{D}_\mathrm{r}^{jk}}.
\label{eq:E_comoving}
\end{equation}
The comoving radiative flux is obtained as follows,
\begin{equation}
\bar{F}_\mathrm{r}^i=
-\frac{1}{1+\Delta\tau_\mathrm{as}\Gamma^{-2}}
\left[\beta_j\bar{D}_\mathrm{r}^{ij}\bar{E}_\mathrm{r}
-\frac{1}{\Gamma}\left(\delta^i_j+\frac{\Gamma^2}{\Gamma+1}\beta^i\beta_j\right){F}_\mathrm{r}^{j,n}
+\beta^iE_\mathrm{r}^n
\right]
\label{eq:F_comoving}
\end{equation}
In other words, for a given set of variables at the previous step $t=t^n$, the comoving radiation energy density and radiative flux are expressed in terms of the velocity and the gas temperature, $\beta^i$ and $\bar{T}_\mathrm{g}$ at $t=t^n+\Delta t$, which are later determined by iteratively solving non-linear equations. 

One of the advantages of using these solutions for $\bar{E}_\mathrm{r}$ and $\bar{F}_\mathrm{r}^i$ is that they behave well even in highly optically thick regimes. 
For a large absorption opacity $\Delta\tau_\mathrm{as},\Delta\tau_\mathrm{ab}\gg 1$, these comoving values are guaranteed to approach to the following asymptotic values;
\begin{equation}
\bar{E}_\mathrm{r}=a_\mathrm{r}\bar{T}_\mathrm{g}^4,\ \ \ \bar{F}_\mathrm{r}^i=0,
\end{equation}
which are expected in optically thick media. 

\subsection{Coupling with hydrodynamics}
The velocity $\beta^i$ and the gas temperature $\bar{T}_\mathrm{g}$ are determined by the coupling between gas and radiation. 
The discretized hydrodynamics equations for the gas energy and momentum densities are given as follows,
\begin{equation}
M^0-M^{0,n}=-G^0\Delta t,\ \ \ M^i-M^{i,n}=-G^i\Delta t,
\end{equation}
where $M^0$ and $M^i$ are conserved variables;
\begin{equation}
M^0=\left(\bar{\rho}+\bar{E}_\mathrm{g}+\bar{P}_\mathrm{g}\right)\Gamma^2-\bar{P}_\mathrm{g},\ \ \ 
M^i=\left(\bar{\rho}+\bar{E}_\mathrm{g}+\bar{P}_\mathrm{g}\right)\Gamma^2\beta^i.
\end{equation}
 
In a similar way to the equations for $E_\mathrm{r}$ and $F_\mathrm{r}^i$, the following transformation gives the following simplified expression,
\begin{eqnarray}
&&\Gamma(M^{0,n+1}-M^{0,n})-\Gamma\beta_i(M^{i,n+1}-M^{i,n})
\nonumber\\
&&=\frac{\Delta\tau_\mathrm{ab}}{\Gamma}
\frac{\left(1+\Delta\tau_\mathrm{as}\Gamma^{-2}\right)
\left(E_\mathrm{r}^{n}-\beta_iF_\mathrm{r}^{i,n}-a_\mathrm{r}\bar{T}_\mathrm{g}^4\right)
+\beta_i\beta^{i}E_\mathrm{r}^n-\beta_iF_\mathrm{r}^{i,n}+\beta_j\beta_k\bar{D}_\mathrm{r}^{jk}a_\mathrm{r}\bar{T}_\mathrm{g}^4}
{\left(1+\Delta\tau_\mathrm{ab}\Gamma^{-2}\right)\left(1+\Delta\tau_\mathrm{as}\Gamma^{-2}\right)-\bar{D}_\mathrm{r}^{ij}}
\end{eqnarray}
The left-hand side of this equation can further be transformed as follows,
\begin{eqnarray}
&&\bar{\rho}\Gamma\left(\frac{\bar{E}_\mathrm{g}}{\bar{\rho}}-\frac{\bar{E}_\mathrm{g}^n}{\bar{\rho}^n}\right)
-(\Gamma-\Gamma^n)(\bar{\rho}^n+\bar{E}_\mathrm{g}^n)+\Gamma(\beta_i-\beta_i^n)M^{i,n}
\nonumber\\
&&=\frac{\Delta\tau_\mathrm{ab}}{\Gamma}
\frac{\left(1+\Delta\tau_\mathrm{as}\Gamma^{-2}\right)
\left(E_\mathrm{r}^{n}-\beta_iF_\mathrm{r}^{i,n}-a_\mathrm{r}\bar{T}_\mathrm{g}^4\right)
+\beta_i\beta^{i}E_\mathrm{r}^n-\beta_iF_\mathrm{r}^{i,n}+\beta_j\beta_k\bar{D}_\mathrm{r}^{jk}a_\mathrm{r}\bar{T}_\mathrm{g}^4}
{\left(1+\Delta\tau_\mathrm{ab}\Gamma^{-2}\right)\left(1+\Delta\tau_\mathrm{as}\Gamma^{-2}\right)-\bar{D}_\mathrm{r}^{ij}},
\label{eq:diff_0}
\end{eqnarray}
where the 1st term in the left-hand side represent the change in the specific gas energy, or the gas temperature for ideal gas, while the 2nd and 3rd terms are proportional to the velocity changes. 
Similarly, the following relations hold,
\begin{eqnarray}
\Gamma(M^{i,n+1}-M^{i,n})-\Gamma\beta^i(M^{0,n+1}-M^{0,n})&=&\Gamma\beta^i(\bar{P}_\mathrm{g}+M^{0,n})-\Gamma M^{i,n}
\nonumber\\
&=&\Delta\tau_\mathrm{as}\bar{F}_\mathrm{r}^j
\left(\delta^i_j-\frac{\Gamma}{\Gamma+1}\beta^i\beta_j\right).
\label{eq:diff_i}
\end{eqnarray}

Equation (\ref{eq:diff_0}) and (\ref{eq:diff_i}) are solved to find $\beta^i$ and $\bar{T}_\mathrm{g}$ by some iterative methods. 
Specifically, we adopt the following method. 
Equation (\ref{eq:diff_i}) can be expressed in the following way,
\begin{equation}
\beta^i=
\frac{\beta^i(1+\Delta\tau_\mathrm{as}\Gamma^{-2})(M^{0,n}+P_\mathrm{g}^n)
-\Delta\tau_\mathrm{as}\Gamma^{-2}(F_\mathrm{r}^i-\Gamma\beta_jP^{ij}_\mathrm{r})}
{(1+\Delta\tau_\mathrm{as}\Gamma^{-2})(M^{0,n}+P_\mathrm{g}^n)
-\Delta\tau_\mathrm{as}\Gamma^{-2}\left[E_\mathrm{r}^n-\Gamma^2\beta_i\beta_jP_\mathrm{r}^{ij}/(\Gamma+1)\right]}
\label{eq:iteraction_0}
\end{equation}
This is not the exact solution for $\Gamma\beta^i$, because the right-hand side obviously includes $\beta^i$ and $\Gamma$. 
However, we find that updating $\Gamma\beta^i$ iteratively by using this expression combined with Equation (\ref{eq:F_comoving}) works well for finding an approximate solution of $\beta^i$ for a given $\bar{T}_\mathrm{g}$. 
On the other hand, Equation (\ref{eq:diff_0}) is solved for $\bar{T}_\mathrm{g}$ by the standard Newton-Raphson method for fixed $\beta^i$. 
Approximate solutions of $\beta^i$ and $\bar{T}_\mathrm{g}$ are obtained by successively updating these two values. 

\section{Post-process calculations\label{sec:locating_photosphere}}
\subsection{Locating the photosphere}
\begin{figure}
\begin{center}
\includegraphics[scale=0.6]{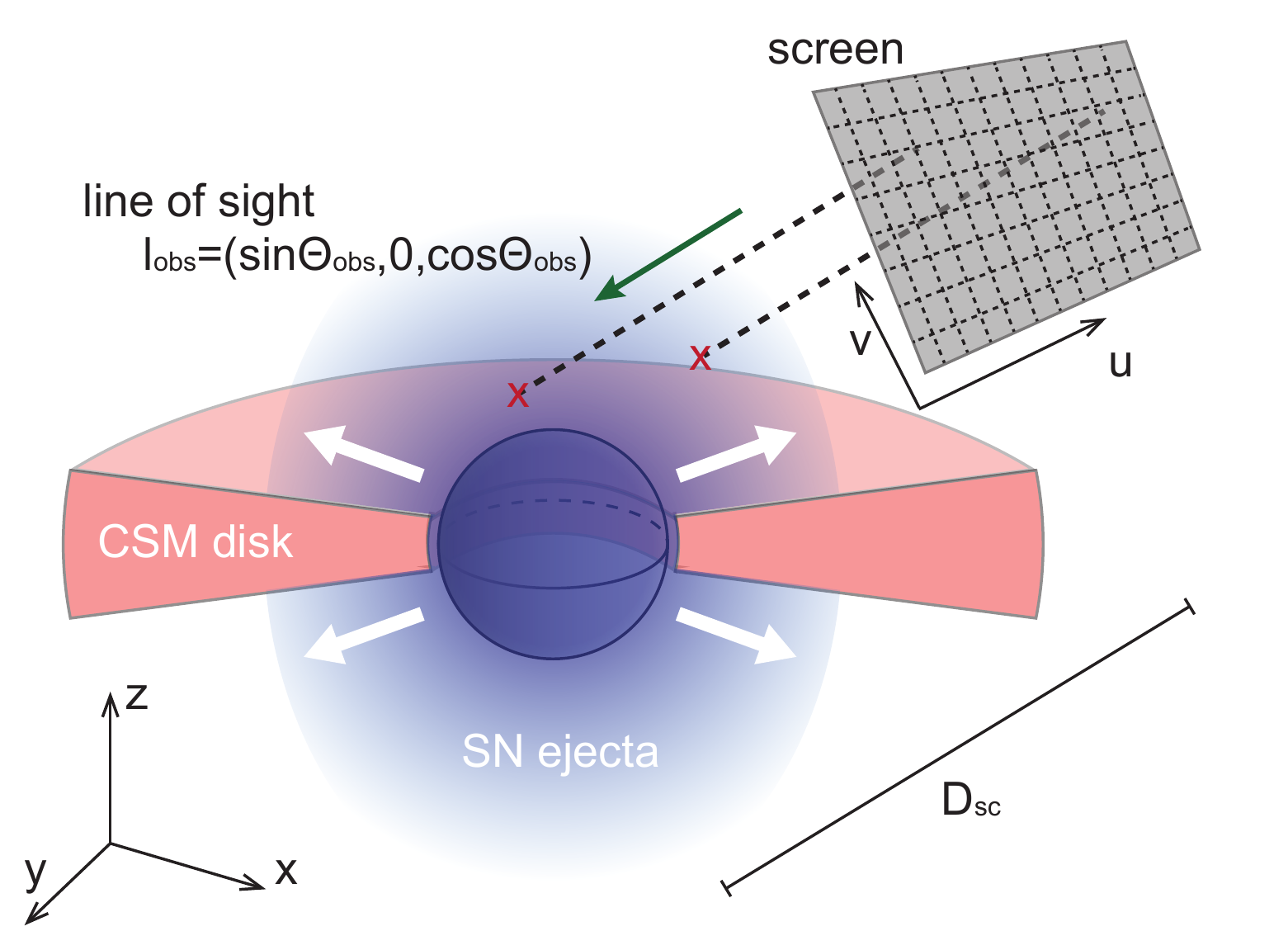}
\caption{Schematic representation of the line of sight integration. 
}
\label{fig:csm_disk}
\end{center}
\end{figure}

First, we map the simulation results into the three-dimensional cartesian space whose origin is identical with that of the simulation coordinate system. 
Then, we consider a rectangle screen extending in the three-dimensional space. 
Figure \ref{fig:csm_disk} schematically represents the situation considered here. 
The screen is extended with an offset $D_\mathrm{sc}$ from the center of the simulation coordinates. 
The two-dimensional cartesian coordinates $(u,v)$ specify a point on the screen. 
The origin of the coordinates is set to be the nearest point to the center of the simulation coordinates. 

We consider photon rays perpendicularly emanating the screen along a viewing angle $\Theta_\mathrm{obs}$. 
Therefore, any point ${\textbf{\textit x}}$ on the ray intersecting with the screen at $(u,v)$ is expressed by introducing a parameter $s$, which specifies the location on the ray;
\begin{equation}
    {\textbf{\textit x}}=u{\textbf{\textit e}}_{u}+v{\textbf{\textit e}}_{v}+s{\textbf{\textit l}}_\mathrm{obs},
\end{equation}
where $\textbf{\textit e}_{u}$ and $\textbf{\textit e}_{v}$ are the unit vectors corresponding to the $u$- and $v$- directions on the screen and $\textbf{\textit l}_\mathrm{obs}=(\sin\Theta_\mathrm{obs},0,\cos\Theta_\mathrm{obs})$ is the direction vector of the ray. 
The screen is located at $s=D_\mathrm{sc}$. 
Along each ray, we integrate the following quantities to obtain the optical depth along the ray,
\begin{equation}
    \tau(u,v,s)=\int _s^{D_\mathrm{sc}}
    \bar{\rho}(\textbf{\textit x})
    (\bar{\kappa}_\mathrm{a}+\bar{\kappa}_\mathrm{s})ds,
    \label{eq:tau}
\end{equation}
where physical variables are taken from a snapshot of the simulations at a specific epoch. The photosphere, beyond which photons can propagate almost freely, is defined so that the optical depth is equal to $2/3$. 
In other words, we find the parameter $s_\mathrm{ph}$ satisfying $\tau(s_\mathrm{ph})=2/3$. 
From the spatial distributions of the radiation energy density and the flux, the (frequency-integrated) intensity at the photosphere along the line of sight is estimated as follows;
\begin{equation}
    I(u,v,s_\mathrm{ph})=\frac{1}{4\pi}\left(cE_\mathrm{r}+3{\textbf{\textit l}}_\mathrm{obs}\cdot{\textbf{\textit F}}_\mathrm{r}\right).
\end{equation}
We assume that the intensity $I(u,v,D_\mathrm{sc})$ at the point $(u,v)$ on the screen is identical with that at the photosphere, $I(u,v,D_\mathrm{sc})=I(u,v,s_\mathrm{ph})$, which holds for radiation propagating in vacuum and is a good approximation for a sufficiently dilute interstellar space. 

In the following, we set the screen distance to be $D_\mathrm{sc}=5\times 10^{16}$ cm, which is identical with $R_\mathrm{obs}$. 
When a snapshot of the simulation at $t=t_\mathrm{sim}$ is used for this post-process calculations, photons emitted from different parts of the photosphere reach the screen at different times. 
The delay time is given by $(D_\mathrm{sc}-s_\mathrm{ph})/c$ and differs from one ray to another according to the difference in $s_\mathrm{ph}$. 
However, dispersion in the delay times, $s_\mathrm{ph}/c$, is generally smaller than the elapsed $t_\mathrm{sim}$, by an order of magnitude. 
Therefore, we neglect the difference in the delay times. 
Then, we use the maximum and minimum delay times, $t_\mathrm{delay,max}=(D_\mathrm{sc}-s_\mathrm{ph,min})/c$ and $t_\mathrm{delay,min}=(D_\mathrm{sc}-s_\mathrm{ph,max})/c$, where $s_\mathrm{ph,min}$ and $s_\mathrm{ph,max}$ are the minimum and maximum coordinates for the locations of the photonsphere among the photon rays, to obtain the average arrival time of photosphereic photons by $t_\mathrm{arrival}=t_\mathrm{sim}+(t_\mathrm{delay,max}+t_\mathrm{delay,min})/2$. 
We assume that the emission properties obtained by this post-process calculation represent those of the emission observed at $t=t_\mathrm{arrival}$. 

\subsection{Color temperature estimation}
The ejecta and the CSM are mostly scattering dominated. 
In other words, photons are predominantly scattered by electrons rather than being absorbed, $\bar{\kappa}_\mathrm{s}>\bar{\kappa}_\mathrm{a}$. 
Therefore, the photosphere identified by the above method is not the region where photons are created and thus the color temperature is determined. 
In order to locate the photon production region, we also calculate the following effective optical depth along the line of sight,
\begin{equation}
    \tau_\mathrm{eff}(u,v,s)=\int _s^{D_\mathrm{sc}}
    \bar{\rho}({\textbf{\textit x}})
    \sqrt{\bar{\kappa}_\mathrm{a}(\bar{\kappa}_\mathrm{a}+\bar{\kappa}_\mathrm{s})}ds',
    \label{eq:tau_eff}
\end{equation}
\citep{1979rpa..book.....R}. 
We locate the effective photosphere so that $\tau_\mathrm{eff}(u,v,s_\mathrm{eff})=2/3$. 
The gas temperature $\bar{T}_\mathrm{g}$ at $(u,v,s_\mathrm{eff})$ gives an estimate for the color temperature $T_\mathrm{c}(u,v)$ of radiation propagating along the line of sight associated with the point $(u,v)$. 

We divide the screen into $1024\times 1024$ cells by equidistant grids and locate the photosphere along the ray associated with each cell. 
As such, we obtain the distribution of the intensity $I(u,v,D_\mathrm{sc})$ and the color temperature $T_\mathrm{c}(u,v)$ on the screen. 
Using these distributions, we calculate the intensity-weighted color temperature, $T_\mathrm{c,av}$, in the following way;
\begin{equation}
    T_\mathrm{c,av}=\frac{\int T_\mathrm{c}(u,v)I(u,v,D_\mathrm{obs})dudv}
    {\int I(u,v,D_\mathrm{obs})dudv}.
\end{equation}
In the above averaging procedure, we only consider rays whose effective optical depth exceeds the threshold of $2/3$ somewhere along the line of sight integration. 
For the given approximated color temperature and the bolometric luminosity $L_\mathrm{bol}$ at the arrival time $t=t_\mathrm{arrival}$, the effective blackbody radius $R_\mathrm{eff}$ is obtained as follows;
\begin{equation}
    R_\mathrm{eff}=\left(\frac{L_\mathrm{bol}}{4\pi \sigma_\mathrm{SB}T_\mathrm{c,av}^4}\right)^{1/2},
    \label{eq:stefan_boltzmann}
\end{equation}
where $\sigma_\mathrm{SB}=5.67\times 10^{-5}\ \mathrm{erg}\ \mathrm{cm}^{-2}\ \mathrm{s}^{-1}\ \mathrm{K}^{-4}$ is the Stefan-Boltzmann constant. 




\bibliography{refs}

\end{document}